\documentclass[a4paper,11pt]{article}
\usepackage[english]{babel}
\usepackage{jheppub}
\pdfoutput=1
\usepackage[T1]{fontenc}
\usepackage{bbold}
\usepackage{float}
\usepackage{amsmath}
\usepackage{empheq}
\usepackage{booktabs}
\usepackage{color}
\usepackage[utf8]{inputenc}
\usepackage{xspace}
\usepackage{scalerel}
\usepackage[most]{tcolorbox}
\usepackage{multirow}
\usepackage{scalefnt}
\usepackage{bold-extra}
\usepackage[shortlabels]{enumitem}
\usepackage[tikz]{bclogo}
\usepackage{subcaption}
\usepackage{tikz-feynman}
\usepackage{cancel}
\usepackage{verbatim}
\usetikzlibrary{arrows,shapes}
\usepackage{afterpage}
\usepackage{graphicx,grffile}
\usepackage{xspace}
\usepackage[normalem]{ulem}
\usepackage{makecell}
\usepackage{mathtools}

\newcommand{\as}{\alpha_s}
\newcommand{\aem}{\alpha_{\rm \scriptscriptstyle{QED}}}

\newcommand{\noun}[1]{{\scshape #1}}

\newcommand{\POWHEG}{\noun{Powheg}}

\newcommand{\POWHEGBOX}{\noun{Powheg-Box}}
\newcommand{\POWHEGBOXRES}{\noun{Powheg-Box-Res}}

\newcommand{\PYTHIA}[1]{\noun{Pythia{#1}}\xspace}
\newcommand{\HERWIG}[1]{\noun{Herwig{#1}}\xspace}

\usepackage{xcolor}

\newtcolorbox{empheqboxed}{colback=white!35, 
 colframe=black,
 width=\textwidth,
 sharpish corners,
 top=-2mm, 
 bottom=0pt
}

\title{
  Resonant leptoquark at NLO with POWHEG}

\author[a]{Luca Buonocore,}
\author[b,c]{Admir Greljo,}
\author[b]{Peter Krack,}
\author[d,e]{Paolo Nason,}
\author[a]{Nud\v zeim Selimovi\'c,}
\author[f]{Francesco Tramontano,}
\author[e,g]{Giulia Zanderighi}

\emailAdd{lbuono@physik.uzh.ch}
\emailAdd{greljo@itp.unibe.ch}
\emailAdd{peter.krack@aiub.unibe.ch}
\emailAdd{paolo.nason@mib.infn.it}
\emailAdd{nudzeim@physik.uzh.ch} 
\emailAdd{francesco.tramontano@na.infn.it}
\emailAdd{zanderi@mpp.mpg.de}

\affiliation[a]{Physik Institut, Universit\"at Z\"urich, CH-8057 Z\"urich, Switzerland}
\affiliation[b]{Albert Einstein Center for Fundamental Physics, Institut f\"{u}r Theoretische Physik, Universit\"{a}t Bern, Sidlerstrasse 5, CH-3012 Bern, Switzerland.}
\affiliation[c]{Department of Physics, University of Basel, Klingelbergstrasse 82, CH-4056 Basel, Switzerland.}
\affiliation[d]{INFN, Sezione di Milano\,-\,Bicocca, and
  Universit\`a di Milano\,-\,Bicocca, Piazza della Scienza 3, 20126 Milano, Italy}
\affiliation[e]{Max-Planck-Institut f\"ur Physik, F\"ohringer Ring 6,
  80805 M\"unchen, Germany}
\affiliation[f]{Universit\`a di Napoli and INFN — Sezione di Napoli,
Complesso Universitario di Monte Sant’Angelo, Via Cinthia 21, 80126 Napoli, Italy}
\affiliation[g]{Physik-Department, Technische Universit\"at M\"unchen, James-Franck-Strasse 1, 85748 Garching, Germany}

\abstract{Recent progress in calculating lepton density functions inside the proton and simulating lepton showers laid the foundations for precision studies of resonant leptoquark production at hadron colliders. Direct quark-lepton fusion into a leptoquark is a novel production channel at the LHC that has the potential to probe a unique parameter space for large masses and couplings. In this work, we build the first Monte Carlo event generator for a full-fledged simulation of this process at NLO for production, followed by a subsequent decay using the POWHEG method and matching to the parton showers utilizing HERWIG. The code can handle all scalar leptoquark models with renormalisable quark-lepton interactions. We then comprehensively study the differential distributions, including higher-order effects, and asses the corresponding theoretical uncertainties. We also quantify the impact of the improved predictions on the projected (HL-)LHC sensitivities and initiate the first exploration of the potential at the FCC-hh. Our work paves the way toward performing LHC searches using this channel.}


\preprint{
  \begin{flushright}  
    MPP-2022-111 \\
    ZU-TH 45/22
  \end{flushright}
}
\begin{document}

\maketitle

\newpage

\section{Introduction}
\label{sec:intro}

Leptoquarks are hypothetical spin-$0$ or spin-$1$ particles carrying both lepton and baryon numbers and mediating a novel interaction among quarks and leptons. They stem from various theories beyond the Standard Model (SM), motivated by the idea of quark-lepton unification. They are predicted in scenarios of matter unification à la Pati-Salam~\cite{Pati:1974yy} involving $SU(4)$ gauge group or in grand unification theories with larger gauge groups like $SU(5)$~\cite{Georgi:1974sy}, and $SO(10)$~\cite{Fritzsch:1974nn}. The phenomenology of leptoquarks is a mature topic (for a recent review see~\cite{Dorsner:2016wpm}), relevant for both low- and high-energy experiments. Leptoquarks at the TeV scale are particularly interesting for high-energy colliders. 

The mass range accessible at colliders is motivated by several extensions of the SM. Scalar leptoquarks arise as pseudo-Nambu-Goldstone bosons of a new strongly interacting sector possibly stabilising the electroweak scale and solving the Higgs hierarchy problem~\cite{Gripaios:2009dq,Fuentes-Martin:2020bnh,Barbieri:2017tuq,Sannino:2017utc,Marzocca:2018wcf}, and are also present in R-parity violating supersymmetric settings~\cite{Giudice:1997wb,Csaki:2011ge,Altmannshofer:2020axr,Dreiner:2021ext}. Vector leptoquarks at the TeV scale are predicted in the partial unification models based on the $SU(4)$ gauge group~\cite{DiLuzio:2017vat,Bordone:2017bld,Greljo:2018tuh,Fornal:2018dqn,Heeck:2018ntp,Cornella:2019hct,Blanke:2018sro,Balaji:2019kwe}. Indirectly, the presence of leptoquarks would impact the low-energy flavour transitions, electroweak precision observables, and Higgs physics. At hadron colliders, a leptoquark would be identifiable as a resonance in the invariant mass of a lepton plus a jet system.

The renewed interest in TeV-scale leptoquarks in recent years originates from several experimental anomalies in semileptonic decays of $B$-mesons~\cite{Lees:2013uzd, Hirose:2016wfn, Aaij:2015yra, Aaij:2014ora, Aaij:2017vbb, Aaij:2013qta, Aaij:2015oid, Aaij:2019wad} which naturally highlight leptoquarks as possible explanation candidates. This is because leptoquarks contribute to the semileptonic transitions at the tree level. At the same time, they affect dangerous four-quark or four-lepton flavour-changing neutral currents, well described by the SM, at the one-loop level only. With anomalies continuing to persist, the TeV-scale leptoquarks with $\mathcal{O}(1)$ couplings to the SM fermions are a clear target for current and future collider searches. Motivated in part by the developments in flavour physics, there has been an increasing effort within ATLAS and CMS experiments to hunt for leptoquarks (for recent results see~\cite{ATLAS:2021jyv,ATLAS:2021oiz,ATLAS:2019qpq,ATLAS:2020xov,ATLAS:2020dsk,CMS:2020wzx,CMS:2018oaj,CMS:2018qqq,CMS:2022nty,CMS:2021far}).

Since leptoquarks couple quarks and leptons, they necessarily carry $SU(3)$ charge and can thus be pair produced in gluon fusion~\cite{Blumlein:1996qp,Kramer:1997hh,Kramer:2004df,Diaz:2017lit,Borschensky:2020hot,Allanach:2019zfr,Borschensky:2022xsa,Dorsner:2018ynv}. This production mechanism is dominant for small values of the leptoquark to quark and lepton coupling $y_{q\ell}$, depending only on the leptoquark mass and the strong coupling $\alpha_s$. However, the pair production is not optimal for heavy leptoquark searches due to rapid phase-space suppression with increasing leptoquark mass. For this reason, often discussed in the literature is the single leptoquark plus lepton production from quark-gluon scattering~\cite{Alves:2002tj,Hammett:2015sea,Mandal:2015vfa,Dorsner:2018ynv}. The production cross section for this process is proportional to $|y_{q\ell}|^2$, but suffers less phase-space suppression, and for $\mathcal{O}(1)$ coupling, it compares to, and even wins over, the cross section for leptoquark pair production. Finally, a non-resonant effect of leptoquarks in the $t$-channel Drell-Yan process could be seen as a deviation in the high-$p_T$ tail of the dilepton invariant mass distribution~\cite{Faroughy:2016osc,Greljo:2017vvb,Schmaltz:2018nls,Fuentes-Martin:2020lea,Greljo:2018tzh,Marzocca:2020ueu,Baker:2019sli,Allwicher:2022gkm}. Since, in this case, the cross-section scales as $|y_{q\ell}|^4$, the expectation is that this process dominates the leptoquark signatures for large couplings and masses beyond the kinematical reach for on-shell production. 

Resonant leptoquark production from a direct lepton-quark fusion is another relevant process at a hadron collider put forward in \cite{Ohnemus:1994xf}. However, before the precise determination of leptonic parton distribution functions (PDF) inside the proton in~\cite{Buonocore:2020nai}, this process could not be utilized in practice. The work of~\cite{Buonocore:2020nai}, therefore, provides a novel opportunity to discover leptoquarks at the LHC. The production cross section scales as $|y_{q\ell}|^2$ and enjoys the least phase-space suppression, making it the most sensitive process for certain parameter regions with $\mathcal{O}(1)$ couplings and TeV-scale masses. The phenomenological collider simulation for this channel, performed in~\cite{Buonocore:2020erb,Haisch:2020xjd}, confirmed this statement and established the resonant leptoquark production mechanism as an exciting candidate for future experimental analyses. In particular, this mechanism surpasses the single leptoquark plus lepton production, which also scales as $|y_{q\ell}|^2$.  

However, the limitation of~\cite{Buonocore:2020erb,Haisch:2020xjd} is the inadequate signal modeling, more precisely, the tree-level approximation and the absence of a lepton shower not available at the time. In this context, the recent computation of next-to-leading (NLO) corrections in~\cite{Greljo:2020tgv}, and the development of a lepton shower in~\cite{Bewick:2021nhc}, constitute a first step toward exploiting the full potential of the resonant leptoquark production, allowing for more advanced precision studies.

In this paper, we make a significant leap forward by constructing a full-fledged Monte Carlo tool for the resonant leptoquark production from a lepton-quark fusion at NLO merged with QCD and lepton showers. By NLO, we mean here that we include QCD corrections and a subset of enhanced QED corrections that makes them comparable with the QCD ones, as we will explain in due time. We quantify the impact of higher-order corrections on the differential phase-space distributions. We assess the importance of the improved signal modeling on the projected LHC bounds derived in~\cite{Buonocore:2020erb}. Additionally, we compute the inclusive cross sections for 100\,TeV proton-proton center-of-mass energy and briefly discuss the potential offered by a future circular hadron collider (FCC-hh)~\cite{FCC:2018vvp}. Our ready-to-use Monte Carlo event generator will facilitate further phenomenological studies and enable the first experimental searches for this process at the LHC.

The paper is organised as follows. In Section~\ref{sec:implementation}, we describe the implementation of the resonant leptoquark production in the \POWHEG{} framework. In Section~\ref{sec:setup}, we define the scope of the leptoquark models and present the expressions for the required amplitudes at NLO. In Section~\ref{sec:decay}, we discuss the treatment of the total decay width, while in Section~\ref{sec:modifications} we report on the modifications of the \POWHEGBOX{} needed to support this process. In Section~\ref{sec:pheno}, we discuss the most important phenomenological imprints of the resonant leptoquark production mechanism. In Section~\ref{sec:xsec}, we validate the implementation against the inclusive NLO cross sections reported in~\cite{Greljo:2020tgv} and provide new results for 100\,TeV collider. In Section~\ref{sec:diffdis}, with quantify the impact of NLO corrections and parton shower effects on the differential distributions. In Section~\ref{sec:sensitivity}, we quantify the error made due to the limited signal simulation in the sensitivity study of~\cite{Buonocore:2020erb}. In Section~\ref{sec:S3example}, we showcase a model example, the $S_3$ leptoquark. In particular, we estimate the sensitivity reach of 100\,TeV proton-proton collider in the mass versus coupling plane. We finally conclude in Section~\ref{sec:conclu}.

\section{Implementation within the \POWHEGBOX{}}
\label{sec:implementation}

In this section we develop a Monte Carlo tool for the resonant leptoquark production at NLO (and subsequent decay) capable of generating Les Houches events (LHE) that can be directly processed by a Parton Shower (PS) program in order to obtain a complete simulation of the collision at the NLO+PS level.

\subsection{Leptoquark models and scattering amplitudes}
\label{sec:setup}

Our goal is a \POWHEGBOXRES{} implementation of the resonant leptoquark production at NLO in both QCD and QED for all renormalisable scalar leptoquark models. The starting point is the Lagrangian in the broken phase, i.e.~respecting $SU(3)_{\rm C} \times U(1)_{\rm QED}$ gauge symmetry,
\begin{equation}\label{eq:Lagdef}
    \mathcal{L} \supset -~y^L_{q\ell} ~\bar q P_{L} \ell ~S_{Q_{\rm LQ}} -~y^R_{q\ell} ~\bar q P_{R} \ell ~S_{Q_{\rm LQ}} ~+~{\rm h.c.}\,,
\end{equation}
where $y^{L,R}_{q\ell}$ are general $3\times3$ Yukawa coupling matrices in flavour space, and $P_{L,R} = (1 \mp \gamma^5) / 2$ are the chiral projectors. The SM chiral fermions $q_{L,R}$ and $\ell_{L,R}$ correspond to the mass eigenstates after the electroweak symmetry breaking. In general terms, the scalar leptoquarks, $S_{Q_{\rm LQ}}$, are triplets of $SU(3)_{\rm C}$, with their possible $U(1)_{\rm QED}$ charges being $|Q_{\rm LQ}|=\left\{\frac{1}{3}, \frac{2}{3}, \frac{4}{3}, \frac{5}{3}\right\}$. The viable flavour structure of the Born process involves the quark flavours $u,d,s,c,b$ (and $t$ at FCC-hh) and charged leptons $e,\mu,\tau$. For instance, a $c$-quark and a $\tau$-lepton can be used to create $S_{1/3}$ and $S_{5/3}$ leptoquarks, as specified by $\mathcal{L} \supset-~y^L_{c\tau} ~ \overline{c_R} \ \tau_L~S_{5/3} -~y^R_{c\tau} ~ \overline{c_L} \ \tau_L^C~S^\dagger_{1/3}\,$. Note that any renormalisable scalar leptoquark model defined respecting the full SM gauge symmetry~\cite{Dorsner:2016wpm} can be recast in the form of Eq.~\eqref{eq:Lagdef} by separately considering the different $SU(2)_{\rm L}$ components of the leptoquark representation. In Section~\ref{sec:S3example}, we will exemplify this with a weak triplet.

Next, we present the expressions for the amplitudes that are used by the code. Starting with the leading order (Born level), the resonant leptoquark production proceeds via fusion of a lepton and a quark in the initial state. Given the energies of the colliders in consideration, we approximate the aforementioned fermions to be massless, such that the possible interference terms between left- and right-handed Yukawa couplings vanish. Thus, the averaged squared matrix element for a particular flavour combination $q\ell$ at the Born level reads
\begin{align}
\texttt{born} &= \frac{1}{4}\left(|y_{q\ell}^L|^2+|y_{q\ell}^R|^2\right) \hat{s}\equiv \frac{1}{4}|y_{q\ell}|^2\, \hat{s}\,,
\end{align}
where $\sqrt{\hat{s}}$ is the partonic-level center of mass energy. Moreover, when different flavour combinations in the initial state contribute to the production of the same leptoquark, the contributions to the averaged squared matrix element are added separately. The expression \texttt{born} matches the input of the \POWHEGBOXRES{} that the process-specific code should provide in the routine
\newline
\phantom{aaaaaa}\texttt{setborn(p(0:3,1:nlegborn), bflav(1:nlegborn), born,\\
\phantom{aaaaaaaaaaaaaa}bornjk(1:nlegborn,1:nlegborn), bmunu(0:3,0:3,1:nlegborn))}~.
\newline
In our case, since we are dealing with the $2\to1$ process, we set $\texttt{nlegborn}=3$, and \texttt{p(0:3,i)} denote the components of the four-momenta of the $\texttt{i}$-th particle. We enumerate the particles in the process $q +\ell \to \rm{LQ}$ as $1 +2 \to 3$, respectively, such that $\hat{s} = \texttt{p(0,3)}^2 -\texttt{p(1,3)}^2 - \texttt{p(2,3)}^2-\texttt{p(3,3)}^2$, while the color correlated squared amplitude, \texttt{bornjk}, and the spin correlated one, \texttt{bmunu}, read
	\begin{align}
			\texttt{bornjk(1,2)} &= \texttt{bornjk(2,3)} = 0\,,\\
			\texttt{bornjk(1,3)} &= \frac{4}{3}\ \texttt{born}\,,\\
			\texttt{bmunu}&=0\,,
	\end{align}
as explained in \cite{Alioli:2010xd}, with $\texttt{bornjk}$ being symmetric. 

Apart from the leading (Born) contribution, this process receives important NLO corrections from interactions with gluons and photons. As shown in~\cite{Greljo:2020tgv}, the QED corrections that we need to include are such that the smallness of the QED coupling is compensated by the PDF enhancement due to the photon in the initial state of the process $\gamma + q \to \ell + \rm{LQ}$, so that they are in fact of the same order as the QCD corrections.

\begin{figure}[t]
\centering
\includegraphics[width=1.0\textwidth]{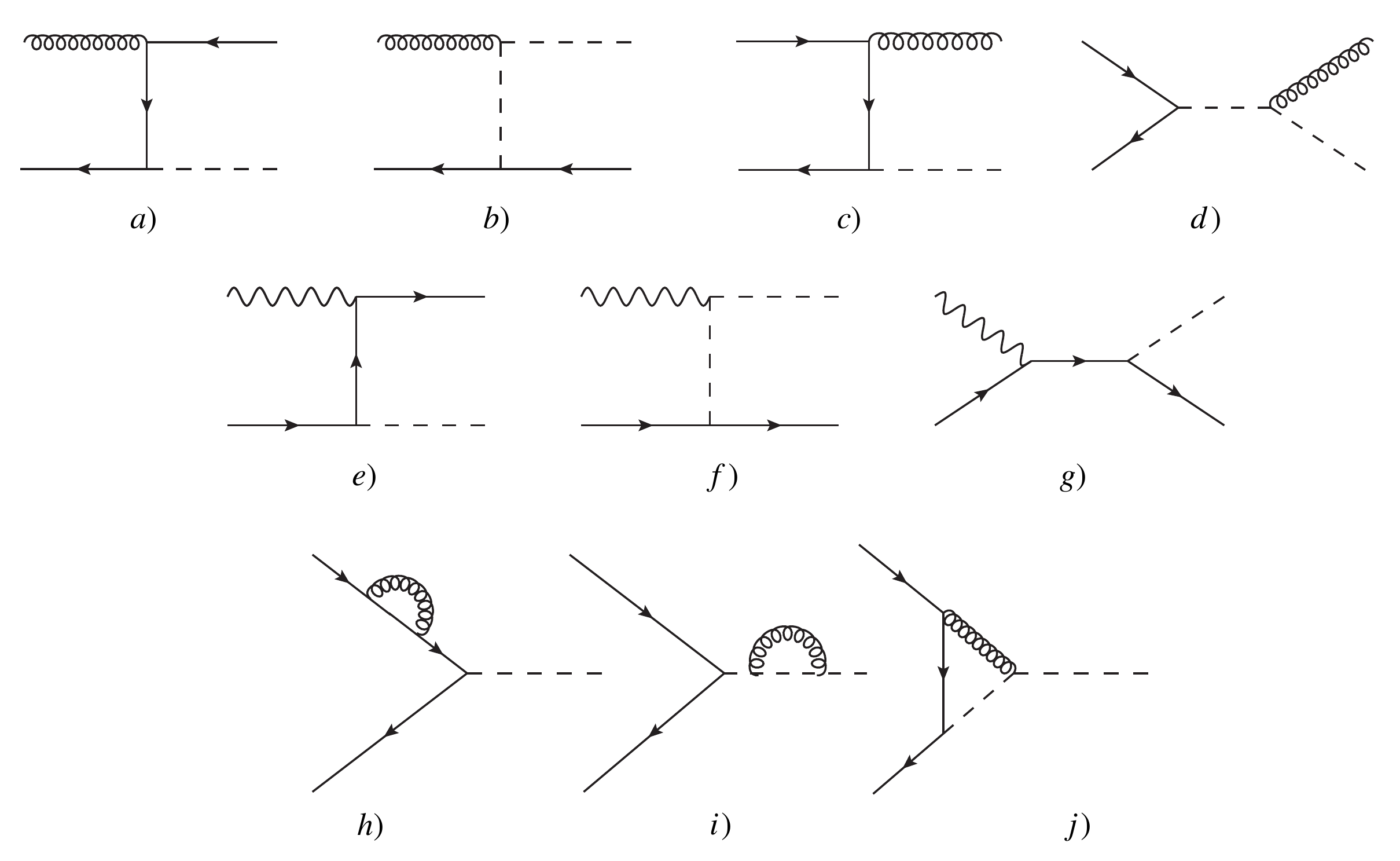}
\caption{Feynman diagrams for the resonant leptoquark production at NLO. Diagrams $a)$ and $b)$ are for $g(p_1) + \ell(p_2)\to q(k) + {\rm{LQ}}(q)$ contributing at $\mathcal{O}(\alpha_s)$. Diagrams $c)$ and $d)$ are for $q(p_1) + \ell(p_2)\to g(k) + {\rm{LQ}}(q)$ contributing at $\mathcal{O}(\alpha_s)$. Diagrams $e)$, $f)$ and $g)$ are for  $\gamma(p_1)+ q(p_2) \to \ell(k) +  {\rm{LQ}}(q)$ contributing at $\mathcal{O}(\alpha)$, but enhanced by the ratio of the photon PDF over the lepton PDF. Finally, diagrams $h)$, $i)$ and $j)$ are the virtual corrections at $\mathcal{O}(\alpha_s)$.}
\label{fig:QCDglue}
\end{figure}

In the context of QCD corrections, the first relevant partonic process is $g(p_1) +\ell(p_2)\to q(k) + {\rm{LQ}}(q)$, shown in Figure~\ref{fig:QCDglue} $a)$ and $b)$. The partonic cross-section for this process was computed in \cite{Kunszt:1997at,Plehn:1997az}.\footnote{Notice that the diagram with the same structure of $b)$, but with an incoming quark and an outgoing lepton, is not included here. It is part of the associated production of a leptoquark and a lepton and should be included in that context. See, for example~\cite{Dorsner:2018ynv}.} Here we present the matrix element squared required by \POWHEGBOXRES{}.  Averaging over spin and colors, and omitting the factor $\alpha_s/2\pi$, with $\alpha_s$ the strong coupling, as specified in \cite{Alioli:2010xd}, it reads
\begin{equation}\label{eq:amp2real1}
\texttt{amp2real}_{g+\ell} = -2\pi^2 |y_{q\ell}|^2 \frac{\hat{s}}{\hat{t}}\,\frac{(\hat{u}^2+m_{\rm{LQ}}^4)}{(\hat{u}-m_{\rm{LQ}}^2)^2}\,,
\end{equation}
where $\hat{s},\hat{t}$, and $\hat{u}$ are the partonic-level Mandelstam 
variables defined as 	
\begin{align}
\nonumber \hat{s} & = (p_1+p_2)^2 = 2\,p_1\cdot p_2 = 2k\cdot q + m_{\rm{LQ}}^2\,,\\
\label{eq:Mandelstam}\hat{t} & = (p_1 - k)^2 = - 2\,p_1\cdot k\,,\\
\nonumber \hat{u} &= (p_1 -q)^2 = -2\,p_1\cdot q + m_{\rm{LQ}}^2\,. 
\end{align}

The second relevant partonic process is with the gluon in the final state $q(p_1) + \ell(p_2)\to g(k) + {\rm{LQ}}(q)$, shown in Figure~\ref{fig:QCDglue} $c)$ and $d)$. Averaging over spin and colors, removing the factor $\alpha_s/2\pi$, the matrix element squared required by \POWHEGBOXRES{} reads
\begin{equation}
\texttt{amp2real}_{q+\ell} =\frac{16}{3}\pi^2|y_{q\ell}|^2\frac{\hat{u}}{\hat{t}}\,\frac{(\hat{s}^2+m_{\rm{LQ}}^4)}{(\hat{s}-m_{\rm{LQ}}^2)^2}~,
\end{equation}
with Mandelstam variables already defined in Eqs.~\eqref{eq:Mandelstam}.

Additionally, in the context of QED corrections, there is an additional real matrix element from diagrams with a photon in the initial state computed in~\cite{Greljo:2020tgv}, $\gamma(p_1) + q(p_2) \to \ell(k)\ + {\rm{LQ}}(q)$, see Fig~\ref{fig:QCDglue} $e)$, $f)$ and $g)$.  The averaged matrix element squared required by \POWHEGBOXRES{}, divided by $\alpha_s/(2\pi)$, reads
\begin{align}
		\label{eq:ampQED}\texttt{amp2real}_{\gamma+q} &=\frac{\aem}{\as}4\pi^2 |y_{q\ell}|^2\left[-Q_\ell^2\frac{\hat{s}}{\hat{t}} - 2 Q_\ell Q_{\rm q} \left(1+\frac{m_{\rm{LQ}}^2 \hat{u}}{\hat{s}\hat{t}}\right)- Q_\ell Q_{\rm LQ}\frac{\hat{u}}{\hat{s}+\hat{t}} \left(1-\frac{2 m_{\rm{LQ}}^2}{\hat{t}}\right)\nonumber\right.\\
		&\left.-Q_{\rm q}^2\frac{\hat{t}}{\hat{s}}+ Q_{\rm LQ}^2\frac{\hat{u}^2 }{(\hat{s}+\hat{t})^2}\left(1+\frac{m_{\rm{LQ}}^2}{\hat{u}}\right)+Q_{\rm q} Q_{\rm LQ}\frac{\hat{u}}{\hat{s}+\hat{t}}\left(1-\frac{2 m_{\rm{LQ}}^2}{\hat{s}}\right)\right]\,,
\end{align}
where $Q_\ell$, $Q_{\rm q}$, and $Q_{\rm LQ}$ are lepton, quark, and leptoquark electric charges, respectively. Since QED preserves charge conjugation, the cross section for the conjugated process is the same, i.e. there are no terms linear in the electric charge in Eq. \eqref{eq:ampQED}. For example, the amplitude is the same  for $\gamma(p_1) +\bar{u}(p_2) \to \ell^-(k) +{\rm LQ}_{1/3}(q)$ and $\gamma(p_1) + u(p_2) \to \ell^+(k) +{\rm LQ}_{-1/3}(q)$. 

Finally, the last ingredient which needs to be provided to \POWHEGBOXRES{} is the finite part of the virtual corrections computed in dimensional regularization. The corresponding diagrams are shown in Figure~\ref{fig:QCDglue} $h)$, $i)$ and $j)$. The result for the finite part of the virtual cross section for the process $q+\ell\to\rm{LQ}$, derived in \cite{Kunszt:1997at}, rewritten in a way to match the form of \texttt{virtual} in \POWHEGBOXRES{} subroutine \texttt{setvirtual} in \cite{Alioli:2010xd} reads
\begin{equation}
	 	\texttt{virtual} = -\frac{|y_{q\ell}|^2}{3} \hat{s} \left[2 + \frac{\pi^2}{6}+\log\left(\frac{\mu_R^2}{\hat{s}}\right)\left(1+\frac{1}{2}\log\left(\frac{\mu_R^2}{\hat{s}}\right)\right)\right]\,,
	 \end{equation}
where $\mu_R$ is the renormalization scale. 

This completes the list of the standard ingredients needed to
set up a process within the \POWHEGBOXRES{} framework. There are
other few peculiar aspects, related to the specific processes at hand,
to be considered. Given that similar issues might also occur for
other BSM applications, we decided to provide a flexible solution
adding and/or modifying some parts of the \POWHEGBOXRES{} code. We
give more details in the following two dedicated sections.

In a full-fledged simulation, the hard scattering process needs to be matched with a parton shower.\footnote{The matching is straightforward in the case of a $p_T$ ordered shower, such as \PYTHIA{8}~\cite{Sjostrand:2014zea}. The matching to an angular ordered shower, such as \HERWIG{}~\cite{Bellm:2019zci}, although more delicate, is also well understood~\cite{Nason:2004rx}.}  Despite the recent interest and progress in the phenomenology of lepton induced processes in hadron-hadron collisions, the availability of Monte Carlo generators which handle initial-state leptons is
rather limited. So far, only \HERWIG{}\cite{Bellm:2019zci} provides a support for showering lepton initiated processes in a development branch which is publicly available~\cite{HW7dev}. As a first
application, a \POWHEG{} NLO+PS generator for various lepton-lepton scattering processes relevant at the LHC has been put forward in~\cite{Buonocore:2021bsf}. In the present work, we make use of a similar setup and refer the interested reader to~\cite{Buonocore:2021bsf} for further details.

\subsection{The line shape and the decay width}
\label{sec:decay}

For most of the parameter space, we can factorize the leptoquark production from decay by using a narrow width (NW) approach. This is a good approximation for inclusive
observables such as the total production rate. However, the relevance of the resonant leptoquark mechanism in the context of the LHC searches crucially relies on modeling the line shape. The leptoquark mass is reconstructed from its decay products, the lepton-jet system. The shape of the reconstructed mass peak depends not only on the intrinsic width but also on QCD and QED radiation, jet reconstruction, and detector resolution. The latter effects, which lead to a broadening of the peak, are usually dominant in weakly-coupled ultraviolet (UV) completions. Therefore, when considering a narrow leptoquark, we can neglect the intrinsic width when generating the fixed order events to be subsequently fed by the parton shower.

However, this approach is not sufficient for UV models featuring moderate-to-strong leptoquark couplings. The description of the leptoquark line shape in our simulation is improved by incorporating the finite width effects. This impacts the large couplings in Eq.~\eqref{eq:Lagdef} predicting a broad resonance, but which are still within the realm of perturbation theory, e.g. $\Gamma_{{\rm LQ}}/m_{{\rm LQ}} \sim \mathcal{O}(0.1)$ where $\Gamma_{LQ}$ is the total leptoquark decay width. Following Ref.~\cite{Gigg:2008yc},
we recast the LO differential partonic cross section for the
production of a scalar leptoquark in lepton-quark collisions in the following
factorised form,
\begin{equation}\label{eq:2to2fact}
 \begin{split}
    d\hat{\sigma}_{{\rm LO}}&= \frac{1}{2s} d\Phi_2  |M_{q+\ell \to q^\prime+\ell^\prime }|^2  \\
    &=  \frac{1}{2M^2}  dM^2 d\hat{\sigma}_{q+\ell \to {\rm LQ}}(M^2)\frac{1}{\pi}\frac{m_{{\rm LQ}}\Gamma_{{\rm LQ}}}{(M^2-m_{{\rm LQ}}^2)^2+m_{{\rm LQ}}^2\Gamma_{{\rm LQ}}^2}\frac{M\Gamma_{{\rm LQ}\to q^\prime+\ell^\prime}(M)}{m_{{\rm LQ}}\Gamma_{{\rm LQ}}}~.
  \end{split}
\end{equation}
Here we used the factorisation properties of the $2\to2$ matrix element, where
\begin{equation}
  d\hat{\sigma}_{q+\ell \to {\rm LQ}}(M^2) =  2\pi\delta(s-M^2)|M_{q+\ell \to LQ}|^2  
  \quad \text{and} \quad
  \Gamma_{{\rm LQ} \to q^\prime+\ell^\prime}(M) = \frac{|y_{q^\prime\ell^\prime}|^2}{16\pi} M ~,
\end{equation}
are the partonic cross section for the production of an on-shell
leptoquark of mass $M$, and the LO partial decay width for ${\rm LQ}
\to q^\prime + \ell^\prime $, respectively.\footnote{We observe that the strict NW approximation is recovered by taking the limit
$
  \frac{1}{\pi}\frac{m_{{\rm LQ}}\Gamma_{{\rm LQ}}}{(M^2-m_{{\rm LQ}}^2)^2+m_{{\rm LQ}}^2\Gamma_{{\rm LQ}}^2}
  \xrightarrow[\Gamma_{{\rm LQ}}\to0]{} \delta(M^2-m_{{\rm LQ}}^2)\,.
$
}

The events are generated according to the following simulation chain:
\begin{enumerate}
\item  Set up \POWHEG{} for a $2\to 1$ kinematic at Born level;
\item  Compute the BRs in the different lepton-quark channels;
\item  Generate isotropic leptoquark decays according to the BRs when
  finalising the LHE events.
\end{enumerate}
Instead, the finite width effects are turned on(off) by setting the flag {\tt  BWgen} to 1(0) in the input card. These effects are included by implementing the last line of Eq.~\eqref{eq:2to2fact} within the above simulation chain as follows:
\begin{enumerate}
\item Set up \POWHEG{} for a $2\to 1$ kinematic at Born level;
\item When generating the Born phase space, add an extra integration over the squared invariant mass $M^2$ (note that the pole mass is $m_{{\rm LQ}}$) in a given finite window $[M^2_{\rm min}, M^2_{\rm max}]$  and include the Breit-Wigner factor as a weight.  That is, generate $M^2$ according to
  \begin{equation}
    dF_{{\rm BW}} = dM^2
    \frac{1}{\pi}\frac{m_{{\rm LQ}}\Gamma_{{\rm LQ}}}{(M^2-m_{{\rm LQ}}^2)^2+m_{{\rm LQ}}^2\Gamma_{{\rm LQ}}^2};
  \end{equation}
\item Compute all Born and real matrix elements setting the leptoquark mass
  to $M$ and include the extra multiplicative factor $f={M^2}/{m_{{\rm LQ}}^2}$ which takes into account the kinematic dependence of the third term in Eq.~\eqref{eq:2to2fact},
  \begin{equation} 
    \frac{M \Gamma_{{\rm LQ}\to q^\prime+\ell^\prime}(M)}{m_{{\rm LQ}} \Gamma_{{\rm LQ}}} = \frac{M^2
      \Gamma_{{\rm LQ}\to q^\prime+\ell^\prime}(M)/M}{m_{{\rm LQ}}^2 \Gamma_{{\rm LQ}}/m_{{\rm LQ}}} =
    \frac{M^2}{m_{{\rm LQ}}^2}{\rm BR}({\rm LQ}\to q^\prime+\ell^\prime)~,
  \end{equation}
\item Compute the BRs for all lepton-quark channels;
\item Generate isotropic leptoquark decays according to the BRs when
  finalising the LHE events.
\end{enumerate}

The partial decay width of the scalar leptoquarks, including the NLO
QCD corrections reads~\cite{Plehn:1997az}
\begin{equation}
	\label{eq:width}
	\Gamma_{{\rm LQ}\to q+\ell} =
        \frac{|y_{q\ell}|^2}{16\pi}m_{\rm{LQ}}\left(1+\frac{\alpha_s}{\pi}\left(\frac{9}{2}-\frac{4\pi^2}{9}\right)\right)\,.
\end{equation}
Here we assume $m_q + m_\ell \ll m_{{\rm LQ}}$. In \POWHEGBOXRES{} the user can opt for using
Eq.~\eqref{eq:width} to compute the total width automatically given the
input coupling matrix, or instead specify the arbitrary value for the width accounting for possibly missing decay channels. 

The expressions for the Born, real, and virtual matrix elements in Section~\ref{sec:setup}, as well as the decay width in Eq.~\eqref{eq:width} are completely general, and are valid for all possible scalar leptoquarks. With the appropriate choice of leptoquark charge $Q_{{\rm LQ}}$ and couplings parametrised by the Yukawa matrices, this code can be used to explore any renormalisable scalar leptoquark model as discussed below Eq.~\eqref{eq:Lagdef}.

\subsection{Modifications of the \POWHEGBOX{}}
\label{sec:modifications}

The \POWHEGBOX{} code provides advanced and automatised
implementation of the FKS subtraction method~\cite{Frixione:1995ms} for computing QCD
and EW corrections in the context of the SM. This allows to add new SM
processes in a straightforward way once the corresponding
matrix elements are available, as described in the previous
sections. However, when dealing with exotic particles for BSM
applications, several process-dependent hacks are required to complete
the calculation.  While working on the implementation of the scalar leptoquark, we
introduced new features necessary for these circumstances.
Therefore, before moving to the phenomenological results, we take the
occasion to document such novelties which are available in
\POWHEGBOXRES{}.\footnote{These modifications are publicly available
for beta testing in the folder Beta-progress.}

The first feature concerns new particles that are charged under the
QCD gauge group, $SU(3)_{\rm C}$.  By default, the program will not recognise
them as a possible emitter of radiation and, correspondingly, will miss to consider
the associated soft singularities.
To account for this, we introduced a facility which
enables the developer of the new process
to assign the colour representation
of the new particle by
calling the subroutine
\begin{verbatim}
                     subroutine set_colour(pid,rep,setget)
\end{verbatim}
The first argument, {\tt integer pid}, is an identification number of
the particle. In principle it should be the identification
  code of the particle according to the Monte Carlo numbering scheme~\cite{Workman:2022ynf}.
  If the particle does not have an identification code any integer value can be used
except for those already
assigned to the SM particles.  The second argument, {\tt character * 4
  rep}, represents the colour representation of the particle. It can
assume the values {\tt '3'},{\tt '3bar'} and {\tt 'adj'} for fundamental,
anti-fundamental and adjoint representations, respectively.  The third
argument, {\tt character * 3 setget} specifies the behavior of the
subroutine. When {\tt setget='set'}, it assigns the representation {\tt
  rep} to the particle with indentifier {\tt pid}. When
{\tt setget='get'}, the subrotuine returns the value of the representation
of the particle with indentifier {\tt pid} and stores it in the
variable {\tt rep}. The latter is required for internal usage, while the process-specific code should just use {\tt 'set'}.  

As an example, our scalar leptoquark transforms according to the fundamental representation of $SU(3)_{\rm C}$. In this case, we just need to
add the following line within the {\tt init\_processes}
\begin{verbatim}
                         call set_colour(42,'3','set')
\end{verbatim}
where we assign the leptoquark with the identifier number $42$. With
this, the program will correctly handle the singular region associated
to the leptoquark emitting a soft gluon including the corresponding soft
terms in the calculation.

The second feature concerns the treatment of the collinear remnants
associated to initial state radiation. \POWHEGBOX{}
automatically generates those contributions on the basis of the
possible underlying Born configurations. The algorithm inspects
initial state partons, and, if they are coloured and/or electrically
charged (when QED corrections are turned on), adds the remnants related
to all possible splitting. However, this mechanism may fail for
non-standard applications. For example, in the scalar leptoquark case, we
consider, together with QCD ones, a subset of QED radiative
corrections associated only to the photon-to-lepton initial-state
splitting. In this case, the algorithm will process an underlying Born amplitude
characterised by the presence of both a lepton and a quark in the
initial state. Since the quark carries a non-vanishing electric
charge, it will also add a spurious remnant associated to the
photon-to-quark emission, which, though possible, is neglected in our
calculation since it is subleading. We overcame this issue by implementing a new version of
the collinear remnants aware of the radiation
regions ({\tt alr} in the nomenclature of \POWHEGBOX{}) that are really present in the calculation,
rather than guessing them according
to the underlying Born configurations. The {\tt alr} are in turn
determined  by the \POWHEGBOX{} based on the real processes specified by the process-specific routines. 

This new mechanism is completely transparent for
the user implementing a new process,
who needs to provide the Born and real
processes only.  In addition, we observe that this new implementation is
also helpful for debugging purposes. Indeed, it makes it easier to
split the calculation in subparts which can be separately tested.

\section{Phenomenology}
\label{sec:pheno}

\subsection{Inclusive cross sections}
\label{sec:xsec}

\begin{table}[ht!]
\centering
{\small
      \begin{tabular}{|c|c|c|c|} 
				\hline 
				$m_\mathrm{LQ}$\,[TeV]& Partons &$\sigma_\mathrm{S^{1/3}}$ [pb]&$\sigma_\mathrm{S^{5/3}}$ [pb]\\ 
				\hline 
				\hline 
				\multirow{6}{*}{ 5.0 } & $u\ +\ e$ & $( 1.06 \times 10^{ -2 } )^{ +3.0 \%}_{ -3.4 \%}\pm 1.5\%$ & $( 1.14 \times 10^{ -2 })^{ +2.9 \%}_{ -3.2 \%}\pm 1.5 \%$ \\\cline{2-4} 

				 &$u\ +\ \mu$ & $( 1.02 \times 10^{ -2 })^{+3.0\%}_{-3.4\%}\pm 1.5\%$ & $(1.1\times 10^{ -2 })^{+2.9\%}_{-3.2\%}\pm 1.5\%$ \\ \cline{2-4} 

				 &$u\ +\ \tau$ & $( 8.72 \times 10^{ -3 })^{+3.1\%}_{-3.5\%}\pm 1.6\%$ & $(9.54\times 10^{ -3 })^{+3.0\%}_{-3.3\%}\pm 1.6\%$ \\ \cline{2-4} 

				 &$c\ +\ e$ & $( 1.48 \times 10^{ -3 })^{+3.7\%}_{-4.2\%}\pm 7.5\%$ & $(1.59\times 10^{ -3 })^{+3.5\%}_{-3.9\%}\pm 7.6\%$ \\ \cline{2-4} 

				 &$c\ +\ {\mu}$ & $( 1.45 \times 10^{ -3 })^{+3.7\%}_{-4.1\%}\pm 7.4\%$ & $(1.55\times 10^{ -3 })^{+3.5\%}_{-3.9\%}\pm 7.5\%$ \\ \cline{2-4} 

				 &$c\ +\ {\tau}$ & $( 1.23 \times 10^{ -3 })^{+3.7\%}_{-4.2\%}\pm 7.5\%$ & $(1.33\times 10^{ -3 })^{+3.5\%}_{-4.0\%}\pm 7.5\%$ \\ \cline{2-4} 

				 &$t\ +\ e$ & $( 3.18 \times 10^{ -4 })^{+7.0\%}_{-7.9\%}\pm 0.6\%$ & $(3.05\times 10^{ -4 })^{+7.3\%}_{-8.3\%}\pm 0.6\%$ \\\cline{2-4} 

				 &$t\ +\ {\mu}$ & $( 3.12 \times 10^{ -4 })^{+7.0\%}_{-7.9\%}\pm 0.6\%$ & $(2.97\times 10^{ -4 })^{+7.3\%}_{-8.3\%}\pm 0.6\%$ \\\cline{2-4} 

				 &$t\ +\ {\tau}$ & $( 2.63 \times 10^{ -4 })^{+7.3\%}_{-8.2\%}\pm 0.6\%$ & $(2.49\times 10^{ -4 })^{+7.6\%}_{-8.6\%}\pm 0.7\%$ \\\cline{2-4} 

				\hline 
				\hline 
				\multirow{6}{*}{ 10.0 } & $u\ +\ e$ & $( 7.28 \times 10^{ -4 } )^{ +2.7 \%}_{ -3.0 \%}\pm 1.8\%$ & $( 7.73 \times 10^{ -4 })^{ +2.6 \%}_{ -2.9 \%}\pm 1.8 \%$ \\\cline{2-4} 

				 &$u\ +\ \mu$ & $( 7.12 \times 10^{ -4 })^{+2.7\%}_{-3.0\%}\pm 1.8\%$ & $(7.57\times 10^{ -4 })^{+2.6\%}_{-2.9\%}\pm 1.8\%$ \\ \cline{2-4} 

				 &$u\ +\ \tau$ & $( 6.15 \times 10^{ -4 })^{+2.8\%}_{-3.1\%}\pm 1.9\%$ & $(6.61\times 10^{ -4 })^{+2.8\%}_{-3.0\%}\pm 1.8\%$ \\ \cline{2-4} 

				 &$c\ +\ e$ & $( 5.25 \times 10^{ -5 })^{+3.3\%}_{-3.7\%}\pm 16.4\%$ & $(5.55\times 10^{ -5 })^{+3.2\%}_{-3.5\%}\pm 16.5\%$ \\ \cline{2-4} 

				 &$c\ +\ {\mu}$ & $( 5.16 \times 10^{ -5 })^{+3.3\%}_{-3.7\%}\pm 16.4\%$ & $(5.46\times 10^{ -5 })^{+3.1\%}_{-3.5\%}\pm 16.4\%$ \\ \cline{2-4} 

				 &$c\ +\ {\tau}$ & $( 4.45 \times 10^{ -5 })^{+3.4\%}_{-3.8\%}\pm 16.3\%$ & $(4.75\times 10^{ -5 })^{+3.2\%}_{-3.6\%}\pm 16.5\%$ \\ \cline{2-4} 

				 &$t\ +\ e$ & $( 1.1 \times 10^{ -5 })^{+6.1\%}_{-6.9\%}\pm 0.7\%$ & $(1.06\times 10^{ -5 })^{+6.3\%}_{-7.1\%}\pm 0.7\%$ \\\cline{2-4} 

				 &$t\ +\ {\mu}$ & $( 1.08 \times 10^{ -5 })^{+6.1\%}_{-6.8\%}\pm 0.7\%$ & $(1.04\times 10^{ -5 })^{+6.3\%}_{-7.1\%}\pm 0.7\%$ \\\cline{2-4} 

				 &$t\ +\ {\tau}$ & $( 9.23 \times 10^{ -6 })^{+6.3\%}_{-7.1\%}\pm 0.7\%$ & $(8.86\times 10^{ -6 })^{+6.5\%}_{-7.3\%}\pm 0.8\%$ \\\cline{2-4} 

				\hline 
				\hline 
				\multirow{6}{*}{ 15.0 } & $u\ +\ e$ & $( 1.14 \times 10^{ -4 } )^{ +2.5 \%}_{ -2.8 \%}\pm 2.1\%$ & $( 1.2 \times 10^{ -4 })^{ +2.4 \%}_{ -2.7 \%}\pm 2.1 \%$ \\\cline{2-4} 

				 &$u\ +\ \mu$ & $( 1.12 \times 10^{ -4 })^{+2.5\%}_{-2.8\%}\pm 2.1\%$ & $(1.18\times 10^{ -4 })^{+2.4\%}_{-2.7\%}\pm 2.1\%$ \\ \cline{2-4} 

				 &$u\ +\ \tau$ & $( 9.77 \times 10^{ -5 })^{+2.8\%}_{-2.9\%}\pm 2.2\%$ & $(1.04\times 10^{ -4 })^{+2.5\%}_{-2.8\%}\pm 2.2\%$ \\ \cline{2-4} 

				 &$c\ +\ e$ & $( 5.44 \times 10^{ -6 })^{+3.0\%}_{-3.3\%}\pm 28.4\%$ & $(5.71\times 10^{ -6 })^{+2.9\%}_{-3.2\%}\pm 28.4\%$ \\ \cline{2-4} 

				 &$c\ +\ {\mu}$ & $( 5.37 \times 10^{ -6 })^{+3.0\%}_{-3.3\%}\pm 28.2\%$ & $(5.64\times 10^{ -6 })^{+2.9\%}_{-3.2\%}\pm 28.3\%$ \\ \cline{2-4} 

				 &$c\ +\ {\tau}$ & $( 4.67 \times 10^{ -6 })^{+3.1\%}_{-3.4\%}\pm 28.2\%$ & $(4.94\times 10^{ -6 })^{+2.9\%}_{-3.3\%}\pm 28.3\%$ \\ \cline{2-4} 

				 &$t\ +\ e$ & $( 9.67 \times 10^{ -7 })^{+5.8\%}_{-6.5\%}\pm 0.9\%$ & $(9.4\times 10^{ -7 })^{+5.9\%}_{-6.6\%}\pm 0.9\%$ \\\cline{2-4} 

				 &$t\ +\ {\mu}$ & $( 9.61 \times 10^{ -7 })^{+5.7\%}_{-6.4\%}\pm 0.9\%$ & $(9.29\times 10^{ -7 })^{+5.9\%}_{-6.6\%}\pm 0.9\%$ \\\cline{2-4} 

				 &$t\ +\ {\tau}$ & $( 8.28 \times 10^{ -7 })^{+5.9\%}_{-6.6\%}\pm 1.0\%$ & $(8.02\times 10^{ -7 })^{+6.1\%}_{-6.9\%}\pm 1.0\%$ \\\cline{2-4} 

				\hline
\end{tabular}   
}
\caption{\label{tab:100up} Inclusive cross sections (in pb) at NLO for the resonant leptoquark production $p p \to {\rm LQ}$ plus $p p \to \overline{{\rm LQ}}$ at $\sqrt{s}=100$\,TeV from up-type quarks and charged leptons. For each flavour combination $q \ell$ reported in the second column, the associated Yukawa coupling in Eq.~\eqref{eq:Lagdef} is $y^L_{q \ell} =1$ while $y^R_{q \ell} =0$. The last two columns are for scalar leptoquarks with electric charges $\pm1/3$ and $\pm5/3$, respectively. The two displayed uncertainties are due to the scale variations (first) and PDF replicas (second). See Section~\ref{sec:xsec} for details.}
\end{table}

\begin{table}[ht!]
\centering
{\small
      \begin{tabular}{|c|c|c|c|} 
				\hline 
				$m_\mathrm{LQ}$\,[TeV]& Partons &$\sigma_\mathrm{S^{2/3}}$ [pb]&$\sigma_\mathrm{S^{4/3}}$ [pb]\\ 
				\hline 
				\hline 
				\multirow{6}{*}{ 5.0 } & $d\ +\ e$ & $( 6.93 \times 10^{ -3 } )^{ +3.0 \%}_{ -3.3 \%}\pm 1.6\%$ & $( 7.18 \times 10^{ -3 })^{ +2.8 \%}_{ -3.2 \%}\pm 1.6 \%$ \\\cline{2-4} 

				 &$d\ +\ \mu$ & $( 6.72 \times 10^{ -3 })^{+2.9\%}_{-3.3\%}\pm 1.6\%$ & $(6.98\times 10^{ -3 })^{+2.9\%}_{-3.2\%}\pm 1.6\%$ \\ \cline{2-4} 

				 &$d\ +\ \tau$ & $( 5.74 \times 10^{ -3 })^{+3.1\%}_{-3.4\%}\pm 1.7\%$ & $(5.99\times 10^{ -3 })^{+3.1\%}_{-3.3\%}\pm 1.7\%$ \\ \cline{2-4} 

				 &$s\ +\ e$ & $( 2.4 \times 10^{ -3 })^{+3.2\%}_{-3.6\%}\pm 3.8\%$ & $(2.48\times 10^{ -3 })^{+3.1\%}_{-3.5\%}\pm 3.9\%$ \\ \cline{2-4} 

				 &$s\ +\ {\mu}$ & $( 2.34 \times 10^{ -3 })^{+3.2\%}_{-3.6\%}\pm 3.8\%$ & $(2.42\times 10^{ -3 })^{+3.1\%}_{-3.5\%}\pm 3.9\%$ \\ \cline{2-4} 

				 &$s\ +\ {\tau}$ & $( 1.99 \times 10^{ -3 })^{+3.3\%}_{-3.7\%}\pm 3.9\%$ & $(2.07\times 10^{ -3 })^{+3.2\%}_{-3.6\%}\pm 3.9\%$ \\ \cline{2-4} 

				 &$b\ +\ e$ & $( 1.11 \times 10^{ -3 })^{+3.8\%}_{-4.3\%}\pm 1.3\%$ & $(1.15\times 10^{ -3 })^{+3.7\%}_{-4.2\%}\pm 1.3\%$ \\\cline{2-4} 

				 &$b\ +\ {\mu}$ & $( 1.09 \times 10^{ -3 })^{+3.8\%}_{-4.3\%}\pm 1.3\%$ & $(1.13\times 10^{ -3 })^{+3.7\%}_{-4.2\%}\pm 1.3\%$ \\\cline{2-4} 

				 &$b\ +\ {\tau}$ & $( 9.22 \times 10^{ -4 })^{+3.9\%}_{-4.5\%}\pm 1.4\%$ & $(9.62\times 10^{ -4 })^{+3.8\%}_{-4.3\%}\pm 1.4\%$ \\\cline{2-4} 

				\hline 
				\hline 
				\multirow{6}{*}{ 10.0 } & $d\ +\ e$ & $( 3.99 \times 10^{ -4 } )^{ +2.7 \%}_{ -2.9 \%}\pm 2.0\%$ & $( 4.11 \times 10^{ -4 })^{ +2.7 \%}_{ -2.8 \%}\pm 2.0 \%$ \\\cline{2-4} 

				 &$d\ +\ \mu$ & $( 3.92 \times 10^{ -4 })^{+2.8\%}_{-2.9\%}\pm 2.0\%$ & $(4.04\times 10^{ -4 })^{+2.7\%}_{-2.8\%}\pm 2.0\%$ \\ \cline{2-4} 

				 &$d\ +\ \tau$ & $( 3.39 \times 10^{ -4 })^{+3.0\%}_{-3.0\%}\pm 2.0\%$ & $(3.51\times 10^{ -4 })^{+2.9\%}_{-2.9\%}\pm 2.0\%$ \\ \cline{2-4} 

				 &$s\ +\ e$ & $( 9.39 \times 10^{ -5 })^{+2.8\%}_{-3.2\%}\pm 6.1\%$ & $(9.64\times 10^{ -5 })^{+2.8\%}_{-3.1\%}\pm 6.1\%$ \\ \cline{2-4} 

				 &$s\ +\ {\mu}$ & $( 9.23 \times 10^{ -5 })^{+2.8\%}_{-3.2\%}\pm 6.1\%$ & $(9.49\times 10^{ -5 })^{+2.9\%}_{-3.1\%}\pm 6.1\%$ \\ \cline{2-4} 

				 &$s\ +\ {\tau}$ & $( 7.98 \times 10^{ -5 })^{+3.0\%}_{-3.2\%}\pm 6.1\%$ & $(8.23\times 10^{ -5 })^{+3.0\%}_{-3.2\%}\pm 6.1\%$ \\ \cline{2-4} 

				 &$b\ +\ e$ & $( 3.62 \times 10^{ -5 })^{+3.4\%}_{-3.9\%}\pm 1.8\%$ & $(3.73\times 10^{ -5 })^{+3.4\%}_{-3.8\%}\pm 1.8\%$ \\\cline{2-4} 

				 &$b\ +\ {\mu}$ & $( 3.57 \times 10^{ -5 })^{+3.4\%}_{-3.9\%}\pm 1.8\%$ & $(3.67\times 10^{ -5 })^{+3.4\%}_{-3.8\%}\pm 1.8\%$ \\\cline{2-4} 

				 &$b\ +\ {\tau}$ & $( 3.08 \times 10^{ -5 })^{+3.5\%}_{-4.0\%}\pm 1.9\%$ & $(3.18\times 10^{ -5 })^{+3.4\%}_{-3.9\%}\pm 1.9\%$ \\\cline{2-4} 

				\hline 
				\hline 
				\multirow{6}{*}{ 15.0 } & $d\ +\ e$ & $( 5.42 \times 10^{ -5 } )^{ +2.6 \%}_{ -2.7 \%}\pm 2.5\%$ & $( 5.56 \times 10^{ -5 })^{ +2.5 \%}_{ -2.6 \%}\pm 2.5 \%$ \\\cline{2-4} 

				 &$d\ +\ \mu$ & $( 5.35 \times 10^{ -5 })^{+2.6\%}_{-2.7\%}\pm 2.5\%$ & $(5.48\times 10^{ -5 })^{+2.6\%}_{-2.6\%}\pm 2.5\%$ \\ \cline{2-4} 

				 &$d\ +\ \tau$ & $( 4.67 \times 10^{ -5 })^{+2.9\%}_{-2.8\%}\pm 2.5\%$ & $(4.8\times 10^{ -5 })^{+2.8\%}_{-2.7\%}\pm 2.5\%$ \\ \cline{2-4} 

				 &$s\ +\ e$ & $( 9.99 \times 10^{ -6 })^{+2.6\%}_{-2.9\%}\pm 10.2\%$ & $(1.02\times 10^{ -5 })^{+2.6\%}_{-2.8\%}\pm 10.2\%$ \\ \cline{2-4} 

				 &$s\ +\ {\mu}$ & $( 9.86 \times 10^{ -6 })^{+2.7\%}_{-2.9\%}\pm 10.2\%$ & $(1.01\times 10^{ -5 })^{+2.6\%}_{-2.8\%}\pm 10.1\%$ \\ \cline{2-4} 

				 &$s\ +\ {\tau}$ & $( 8.59 \times 10^{ -6 })^{+2.7\%}_{-3.0\%}\pm 10.1\%$ & $(8.83\times 10^{ -6 })^{+2.8\%}_{-2.9\%}\pm 10.1\%$ \\ \cline{2-4} 

				 &$b\ +\ e$ & $( 3.2 \times 10^{ -6 })^{+3.2\%}_{-3.6\%}\pm 2.6\%$ & $(3.27\times 10^{ -6 })^{+3.2\%}_{-3.5\%}\pm 2.6\%$ \\\cline{2-4} 

				 &$b\ +\ {\mu}$ & $( 3.16 \times 10^{ -6 })^{+3.2\%}_{-3.6\%}\pm 2.6\%$ & $(3.24\times 10^{ -6 })^{+3.2\%}_{-3.6\%}\pm 2.6\%$ \\\cline{2-4} 

				 &$b\ +\ {\tau}$ & $( 2.75 \times 10^{ -6 })^{+3.3\%}_{-3.7\%}\pm 2.7\%$ & $(2.83\times 10^{ -6 })^{+3.2\%}_{-3.6\%}\pm 2.7\%$ \\\cline{2-4} 

				\hline
\end{tabular} 
}
\caption{\label{tab:100down} Inclusive cross sections (in pb) at NLO for the resonant leptoquark production $p p \to {\rm LQ}$ plus $p p \to \overline{{\rm LQ}}$ at $\sqrt{s}=100$\,TeV from down-type quarks and charged leptons. For each flavour combination $q \ell$ reported in the second column, the associated Yukawa coupling in Eq.~\eqref{eq:Lagdef} is $y^L_{q \ell} =1$ while $y^R_{q \ell} =0$. The last two columns are for scalar leptoquarks with electric charges $\pm2/3$ and $\pm4/3$, respectively. The two displayed uncertainties are due to the scale variations (first) and PDF replicas (second). See Section~\ref{sec:xsec} for details.}
\end{table}

The inclusive NLO cross sections for the resonant leptoquark production at the LHC were first computed in~\cite{Greljo:2020tgv}. We have validated our \POWHEGBOXRES{} implementation against this reference.\footnote{While performing the comparison, we found differences at the level of $\lesssim1\%$ with Tables 1 and 2 of~\cite{Greljo:2020tgv} published by some of us. By further inspection, we could trace these differences to the numerical values of $\aem$, the use of negative PDFs, and a missing term in the numerical implementation of the plus distribution. We do not report the same tables again since the phenomenological relevance of these differences is negligible. The updated numbers can easily be obtained using the \POWHEGBOXRES{} implementation, a publicly available supplement to this paper.}

We then complement the study carried out in~\cite{Greljo:2020tgv}, which focuses on the LHC phenomenology, and compute the production rates at the FCC-hh ($\sqrt{s} = 100$\,TeV). In Tables~\ref{tab:100up} and \ref{tab:100down}, we show the \POWHEGBOXRES{} predictions for the inclusive resonant leptoquark production cross sections at $100\,$TeV proton-proton collider. We sum up the two cross sections for particle and antiparticle production. We consider all possible flavour and charge combinations for three different leptoquark masses, $m_{{\rm LQ}} = 5, 10, {\rm and}\,15$\,TeV.  The couplings are all set to zero but for a single entry in $y_{q\ell}^L$ corresponding to a desired quark-lepton flavour combination. (The same results are obtained for $y_{q\ell}^R$ instead of  $y_{q\ell}^L$.) We also compute theoretical uncertainties associated to missing higher orders by taking the envelop of the costumary seven point scale variations, and the error associated with the uncertainty on the pdf which is derived by calculating the symmetric error obtained by averaging the results for all the different replicas.

We make use of the PDF set {\tt LUXlep-NNPDF31\_nlo\_as\_0118\_luxqed}~\cite{Buonocore:2020nai} which includes photons and leptons. Before discussing the results, some comments are in order. At these energies, top quarks are substantially produced by QCD radiation and should be considered as a possible initial state. This opens an opportunity to study new quark-lepton combinations not present at the LHC. A consistent description of top-initiated processes would require using a PDF set with $n_F=6$ flavours. However, a single PDF set that includes both photon/leptons and the top quark is still unavailable. To circumvent this issue, we assume that the presence of top and photon/leptons induces only a slight modification to the dominant partons through DGLAP evolution and affects QCD sum rules, such as the proton momentum conservation, by a tiny amount. Therefore, as a first approximation, we can borrow the top quark density as is from the {\tt NNPDF31\_nlo\_as\_0118\_nf\_6} set~\cite{NNPDF:2017mvq} and add it to the {\tt LUXlep-NNPDF31\_nlo\_as\_0118\_luxqed} set.\footnote{We have also verified that doing the opposite, namely borrowing the photon and leptons from the {\tt LUXlep-NNPDF31\_nlo\_as\_0118\_luxqed} and adding them to the {\tt NNPDF31\_nlo\_as\_0118\_nf\_6}, leads to minor differences at the percent level. Therefore, the uncertainty associated with the above approximation is relatively small and well within the scale uncertainty. Once the appropriate PDF set becomes available, our calculations can easily be repeated using \POWHEGBOXRES{}.}

We, therefore, treat the top quark as an extra initial light parton. Alternatively, given that the top mass acts as the physical regulator of the collinear divergence, one may compute the top process starting from a gluon splitting $g\to t {\bar t}$, retaining the full mass dependence and taking into account the possibility of having a second resolved top. This is in analogy to the 4FS versus 5FS computations for the processes involving bottom quarks at the LHC.

As for the lepton in the initial state, a similar situation also holds for contributions due to the massive EW gauge bosons, whose relevance grows with the collider energies. One can account for them by including additional subprocesses initiated by an EW gauge boson parton splitting into a lepton pair~\cite{Fornal:2018znf,Bauer:2018arx}. We observe that in this case, one can study leptoquark production in neutrino-quark fusion. The account of these effects is beyond the aim of the present work.

The results obtained in Tables~\ref{tab:100up} and \ref{tab:100down} indicate promising prospects at the FCC-hh given the luminosity target is up to $30\,$ab$^{-1}$~\cite{FCC:2018vvp}. As expected from the PDF, the cross sections for heavier quark generations are hierarchically smaller but comparable for different lepton generations.  As anticipated, the top-induced cross sections are sizeable, offering unique opportunities for leptoquarks exclusively coupled to top quarks.
Given these results, it is interesting to analyze the potential offered by FCC-hh in more detail. We relegate further discussion to Section~\ref{sec:S3example} where we perform a simplified sensitivity study based on the total cross section to chart the parameter space for which one expects to produce more than 100 events.

\subsection{Differential distributions}
\label{sec:diffdis}

\begin{figure}[htb]
\centering
{\includegraphics[width=0.42\columnwidth]
   {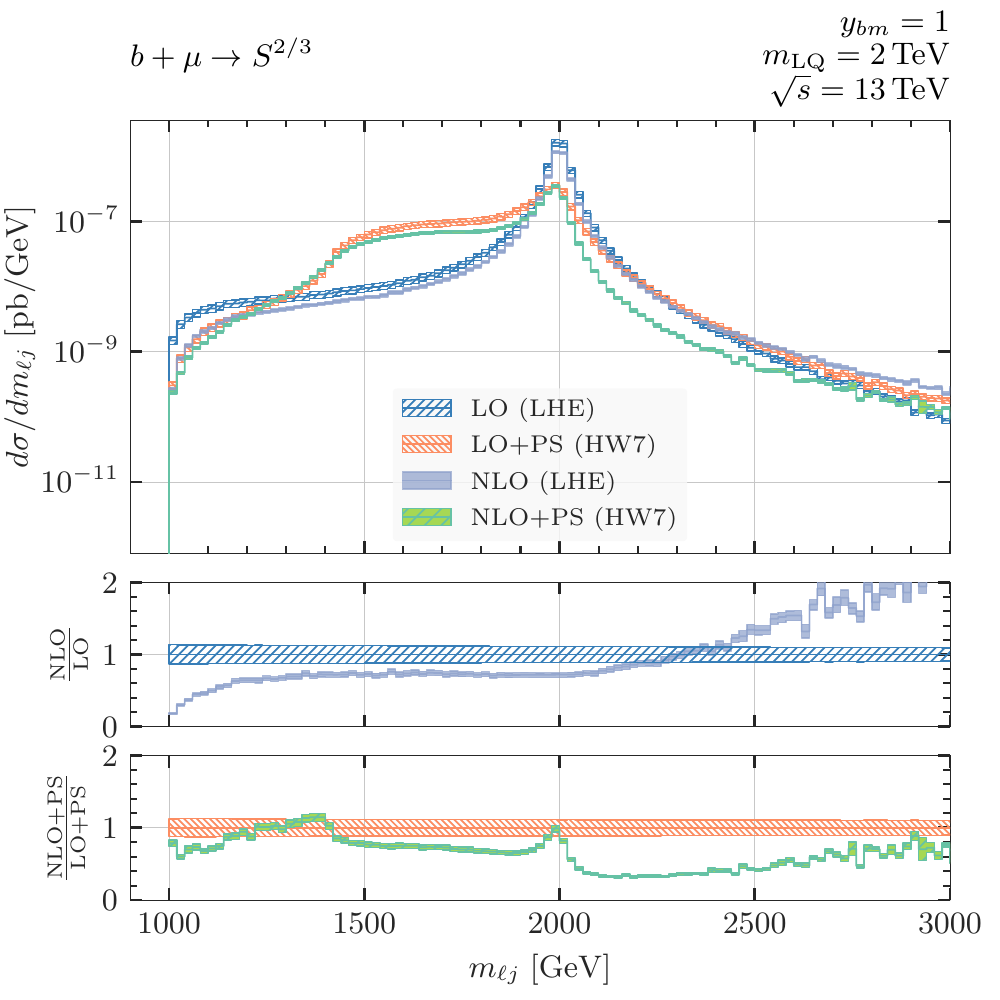}}
{\includegraphics[width=0.42\columnwidth]
   {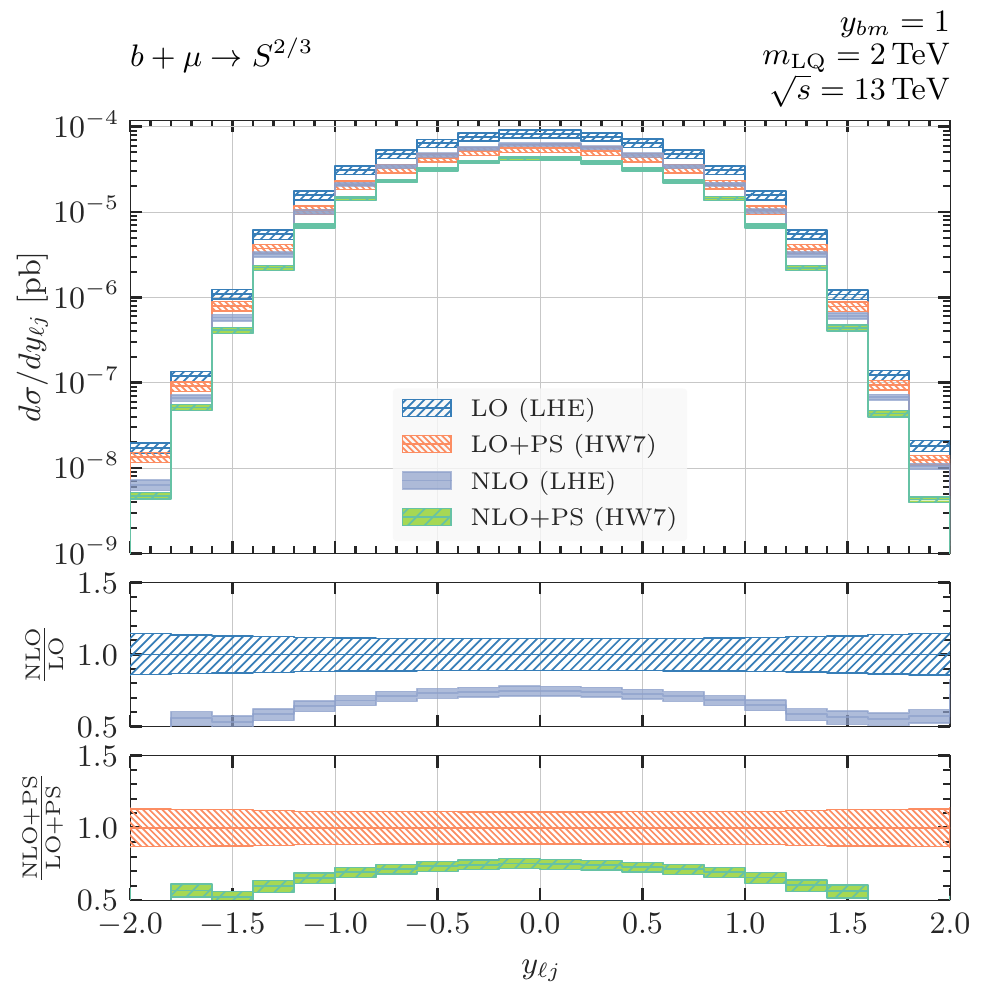}} \\
   \vspace{.1cm}
{\includegraphics[width=0.42\columnwidth]
   {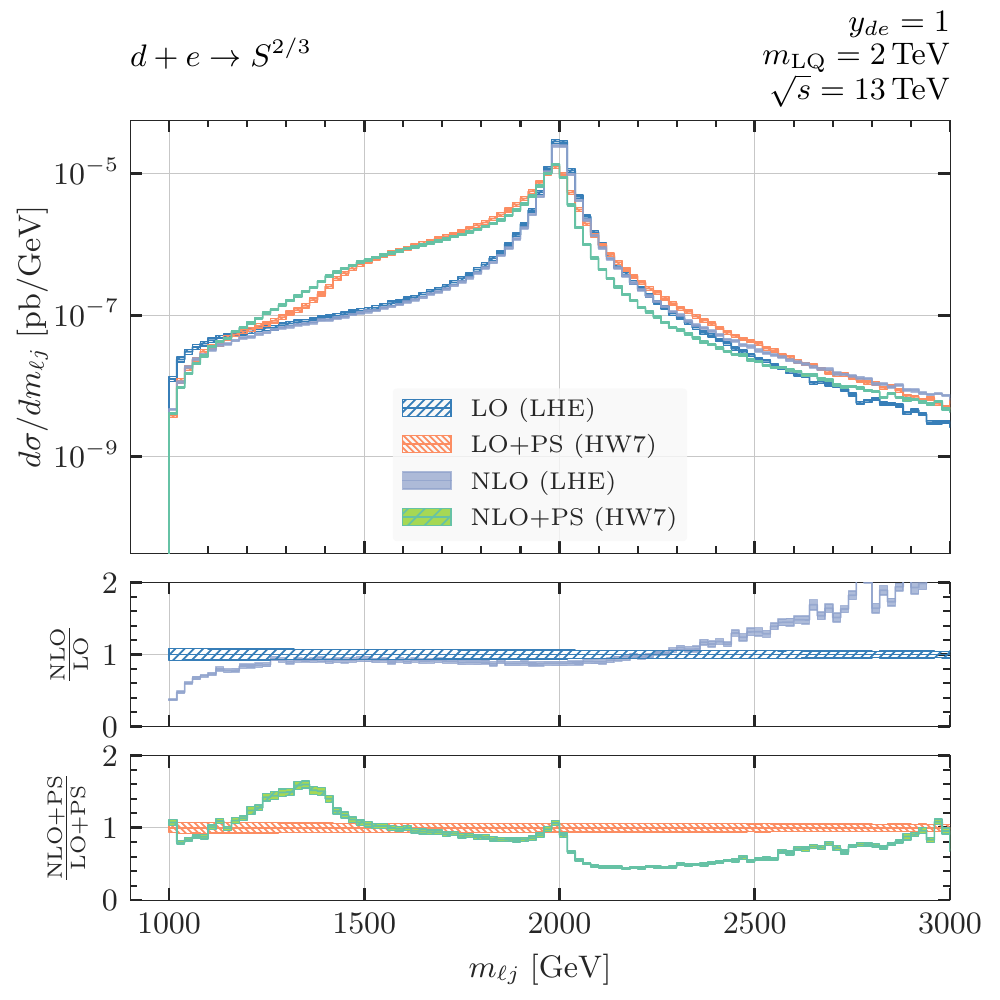}}
{\includegraphics[width=0.42\columnwidth]
   {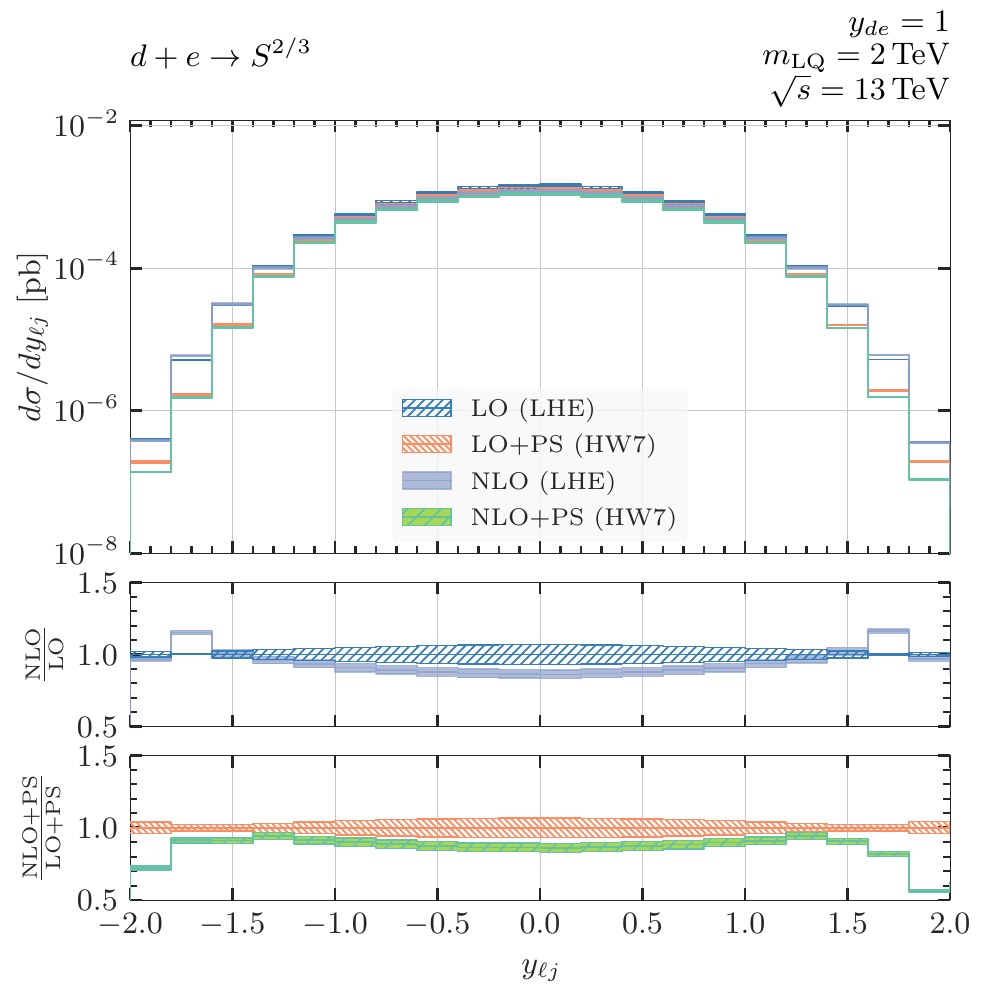}} 
 \caption{{\label{fig:mLQyLQ}} Differential distributions for the benchmark points defined in Section~\ref{sec:diffdis}.  The left (right) panel shows the jet-lepton system's invariant mass (rapidity). }
\end{figure}

\begin{figure}[htb]
\centering
{\includegraphics[width=0.42\columnwidth]
   {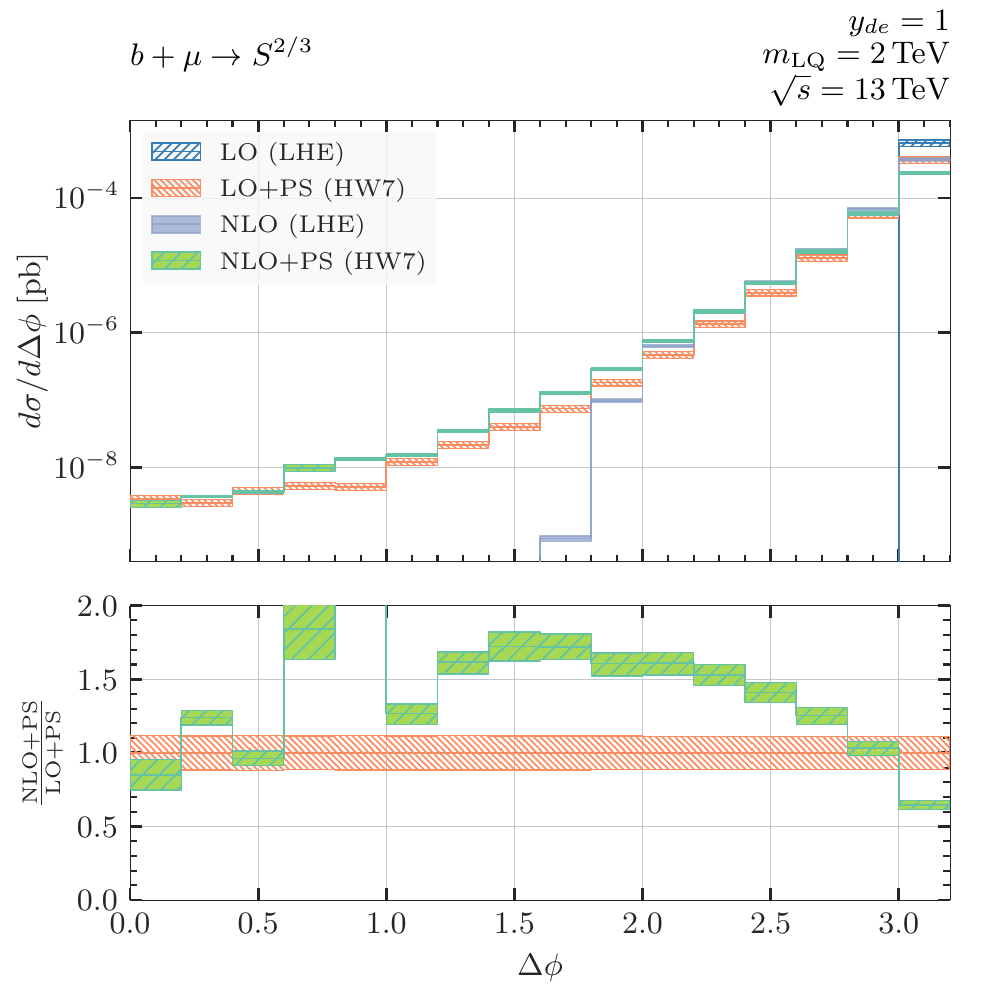}}
{\includegraphics[width=0.42\columnwidth]
   {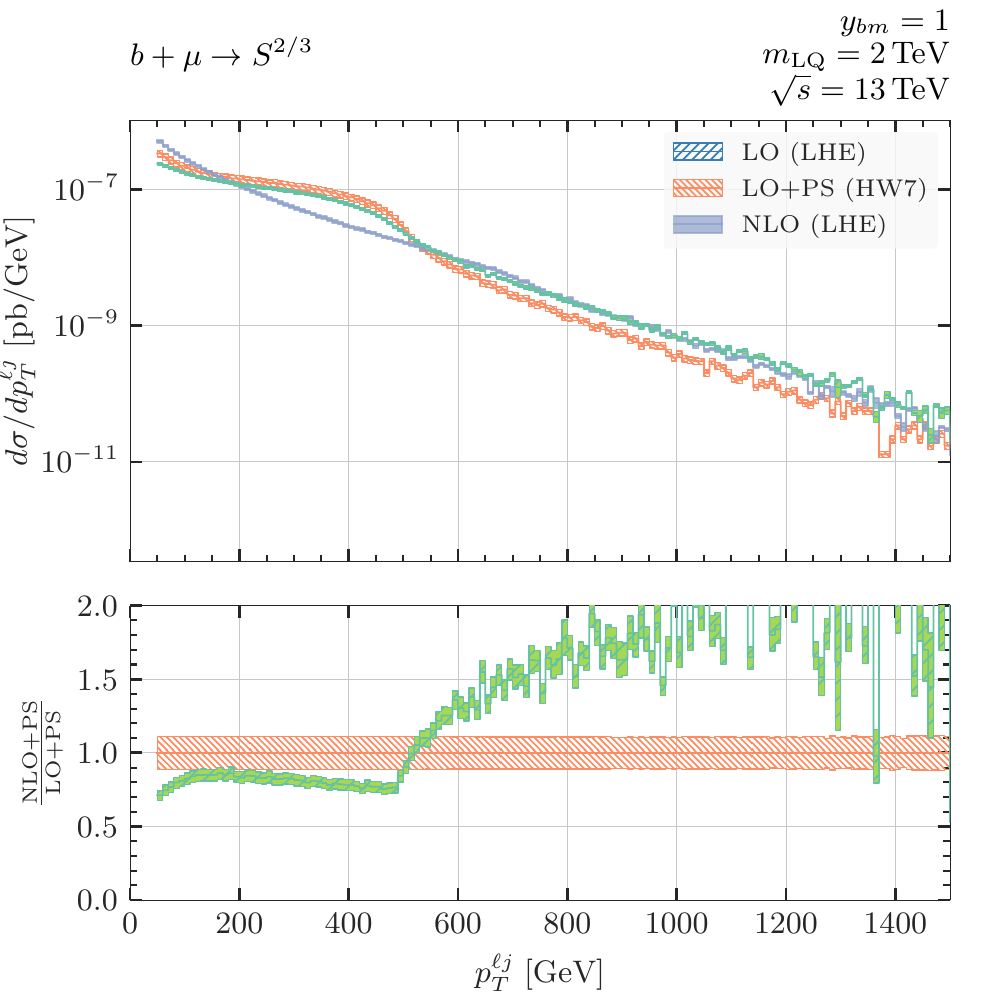}} \\
   \vspace{.1cm}
{\includegraphics[width=0.42\columnwidth]
   {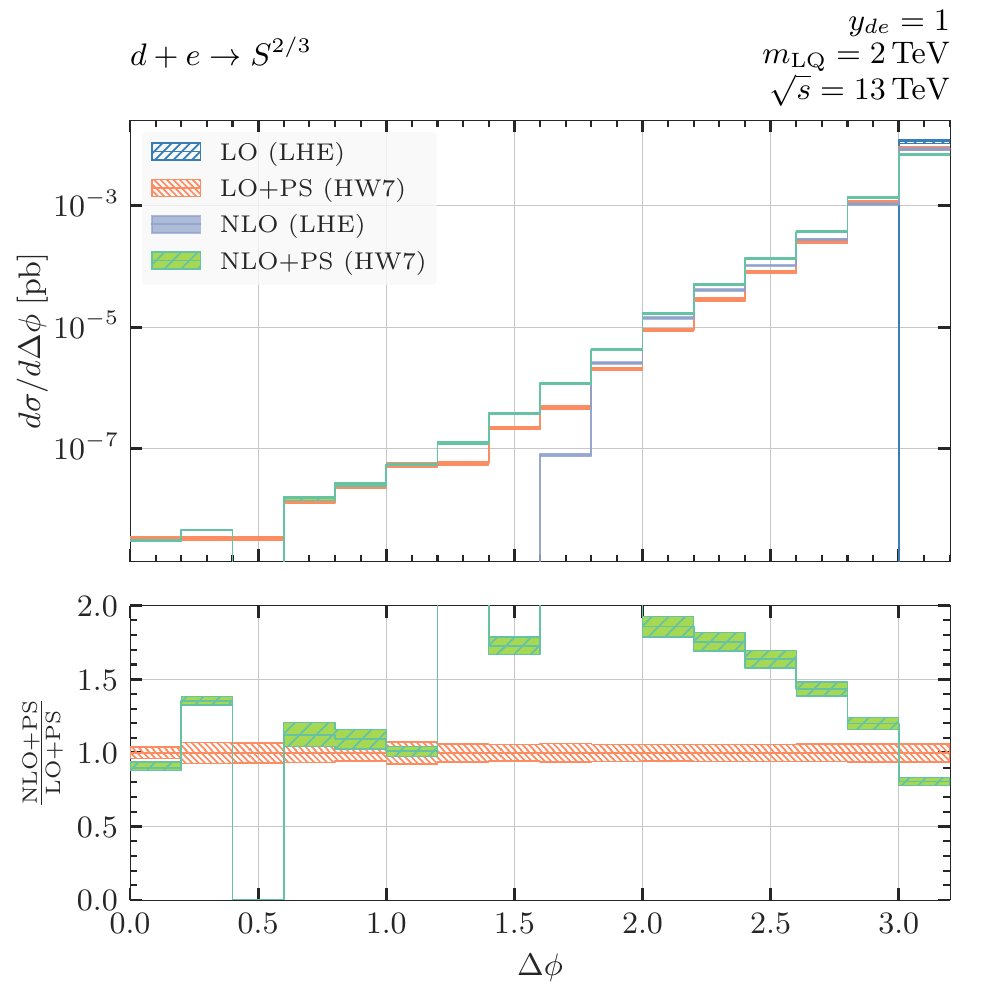}}
{\includegraphics[width=0.42\columnwidth]
   {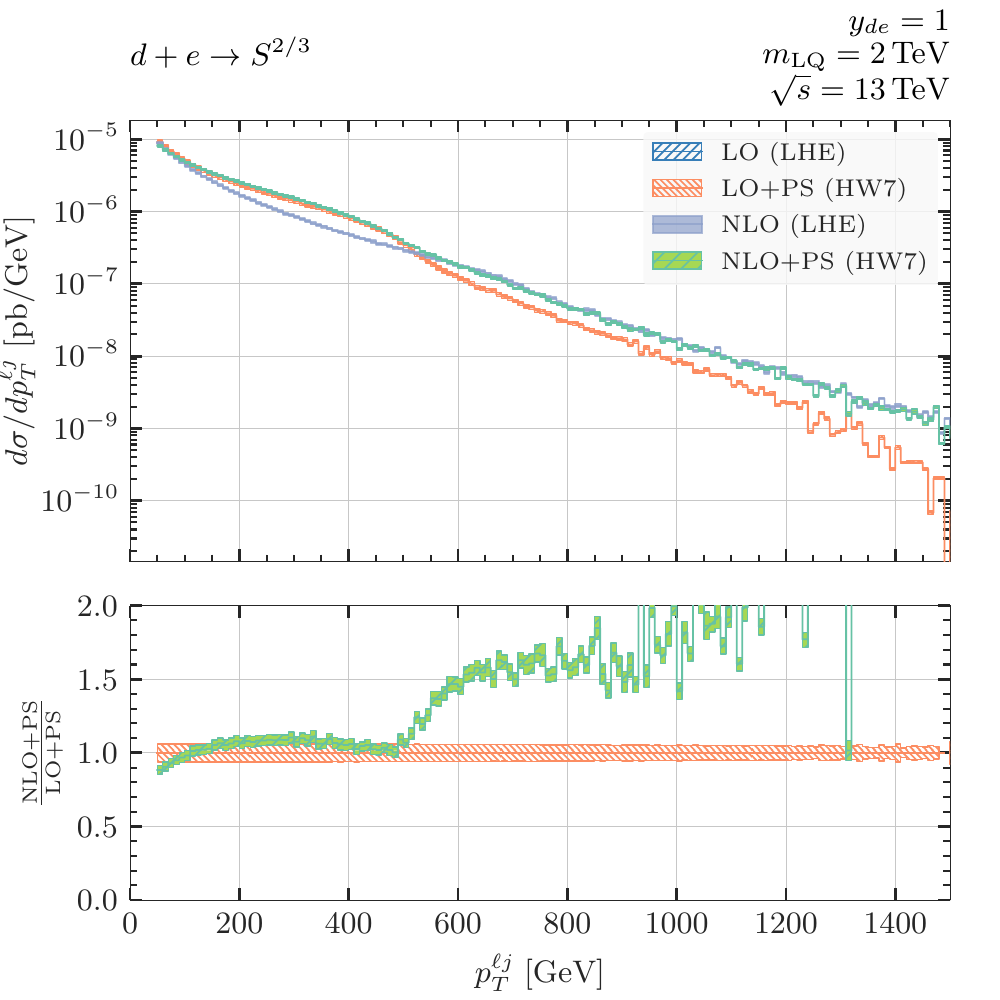}} 
\caption{{\label{fig:dphiptlq}} Differential distributions for the benchmark points defined in Section~\ref{sec:diffdis}. The left panel shows $\Delta
 \phi$ between lepton and jet, and the right panel shows the jet-lepton system's transverse momentum $p_T^{lj}$.}
\end{figure}

The main advantage of the \POWHEGBOXRES{} implementation with respect to~\cite{Greljo:2020tgv} is the flexibility to study arbitrary differential distributions. This can be done at any simulation stage (before and after leptoquark decay or parton shower) at the LO and NLO accuracy. In this section, we comprehensively study the resonant leptoquark production kinematics.

We investigate three benchmark scenarios where the scalar leptoquark is exclusively produced from $b+e$, $b+\mu$, and $d+e$ fusion.\footnote{The code provided in the \POWHEGBOX{} repository~\url{http://powhegbox.mib.infn.it} allows studying other benchmarks efficiently.} In all cases, the leptoquark charge is set to $\pm 2/3$, while the leptoquark mass is set to $m_{{\rm LQ}} = 2$\,TeV for illustration. The Yukawa couplings in Eq.~\eqref{eq:Lagdef} are all set to zero except for the desired quark-lepton flavour combination $y^L_{q\ell}=1$. The leptoquark is therefore decayed to the same quark-lepton pair. The code automatically computes the total leptoquark decay width using Eq.~\eqref{eq:width}. The energy of the proton beams is set to 6.5\,TeV each ($\sqrt{s} = 13$\,TeV) and the PDF set is {\tt{LUXlep-NNPDF31\_nlo\_as\_0118\_luxqed}} (central)~\cite{Buonocore:2020nai}.

When running the reconstruction analysis, a perfect detector is assumed with no smearing effects. The high-level objects of interest are the leading-$p_T$ jet and lepton, which typically originate from the leptoquark decay. The jets are built using the anti-$k_T$ algorithm~\cite{Cacciari:2008gp} with $\Delta R = 0.4$ as implemented in {\tt Fastjet}~\cite{Cacciari:2011ma}. The cuts on the transverse momentum $(p_T > 500\;\mathrm{GeV})$ and the pseudorapidity $(|\eta| < 2.5)$ are applied, and the hardest jet and lepton are then selected.
We also require the total invariant mass of the jet-lepton system to be
above 1~TeV and below 4~TeV.
Bremsstrahlung recombination was considered for the lepton by adding photons that lie inside a cone of $\Delta R < 0.2$. We see no appreciable difference between the muon and the electron case. This is consistent with the fact that
for high-mass objects the quark-electron and quark-muon luminosities are very similar (See Figure~8 in ref.~\cite{Buonocore:2020nai}), and thus
the only substantial difference between muons and electrons is the more significant QED radiation of the latter. So, after recombination, no relevant difference remains. Thus, the $b e$ case is not shown in the figures since it is indistinguishable from the $b \mu$ case. Finite width effects are turned on
(see Sec.~\ref{sec:decay}).

Two million LO and NLO events were processed to generate the plots shown in the Figures~\ref{fig:mLQyLQ} and~\ref{fig:dphiptlq}. The upper box in each plot displays four lines. The bands show the customary 7-point scale variation uncertainty. The lines labeled LO (LHE), and LO+PS (HW7) are obtained by restricting the event generation with \POWHEG{} to leading order and running the analysis before (blue) and after (orange) the parton shower using \HERWIG{}. Similarly, the NLO simulations were used for the purple (before PS) and the green (after PS) lines. The two smaller boxes below the main box show the ratio of NLO to LO before and after showering. As illustrated by the plots, the NLO order corrections are sizeable and depend on the kinematics.

The invariant mass and rapidity of the jet-lepton system are shown in Figure~\ref{fig:mLQyLQ} in the left and right columns, respectively, for the benchmark scenarios: $b+\mu$ (top), and $d+e$ (bottom).
The jet-lepton system's invariant mass distributions ($m_{{\ell j}}$) show a resonance peak at $m_{{\rm LQ}} = 2$\,TeV. The width of the peak before the parton shower is narrow and meets the expectation from the intrinsic leptoquark width in Eq.~\eqref{eq:width}. The effect of the parton shower can be observed as well --- the peak position is shifted to a slightly lower value of ($m_{{\ell j}}$), the peak is smaller and considerably broader for the distributions of showered events both at LO and NLO and leans towards lower masses. It is helpful to remark that this effect is mostly due to the fact that, in our NLO calculation, we do not include radiative corrections to the leptoquark decay. Under these circumstances, the final state radiation generated by the Monte Carlo in the leptoquark decay becomes very relevant since it is the only source of jet momentum degradation due to final state radiation outside the jet cone. This causes a sizeable raise of the backward tail and a slight lowering of the forward tail in the invariant mass spectrum. By investigating further this effect, we have found that another contributing factor is the presence of events such that the selected jet is not the one arising from the leptoquark decay. This is more likely to happen in the showered events, since there are more jets in that case. These events tend to inflate the differential distribution below the peak.

When comparing the jet-lepton system's rapidity plots in Figure~\ref{fig:mLQyLQ} (right column), it is worth noticing a broader distribution for the down quark compared to the bottom quark. This behavior stems from the down quark being a valence quark and having a higher probability of carrying a more significant fraction of the proton's total momentum. Therefore, leptoquarks produced from valence quarks tend to carry more momentum along the beam axis, broadening the shape of the rapidity distribution towards larger values.

Notice also the large difference in the signal rate between the $d$ and the $b$ case, due to the sea versus valence quark PDF, and to larger NLO corrections in the latter. These features are expected since, as shown in Figure~8 of Ref~\cite{Greljo:2020tgv} for the LQ of charge $\pm 2/3$, the NLO QCD K-factor is close to unity for bottom initiated processes and it does not compensate the negative NLO QED one, which is similar for all quark cases.

Figure~\ref{fig:dphiptlq} 
shows the azimuthal angle $\Delta \phi$ between the lepton and the jet in the left column, and the transverse momentum $p_T^{\ell j}$ of the jet-lepton system in the right column. The $\Delta \phi$ between the jet and the lepton in Figure~\ref{fig:dphiptlq} (left column) shows that the two objects are mostly back to back in the azimuthal plane. At LO without parton shower, all events exactly have $\Delta \phi = \pi$. The radiation at NLO opens up smaller angles to the distribution. The parton shower populates even smaller angles, but the rate still clearly peaks around $\pi$ as expected.
The rapid fall of the LHE band in the $\Delta \phi$ plot, near $\delta \phi=1$, can be understood as a kinematic effects. As  $\delta \phi$ decreases, the
transverse momentum of the jet balancing the leptoquark must increase, up to the point when it becomes the hardest jet, and is thus selected as such. The presence of more jets in the shower case can allow instead for a larger boost of the leptoquark, not associated with a single hard jet in acceptance.

The transverse momentum of the jet-lepton system $p_T^{\ell j}$ in Figure \ref{fig:dphiptlq} (right column) is zero for all LO events before the shower.
We notice the feature of the distribution for transverse  momenta between 200 and 500 GeV, where the showered
events have larger cross section, and smaller cross section above 500 GeV. First of all, we have verified that
such feature is not present in the distribution of the leptoquark at the ``Monte Carlo Truth'' level (i.e.
the leptoquark in the Monte Carlo just before decay), where
a perfect agreement is found between the NLO(LHE) and NLO+PS(HW7) distributions. This is due to the fact that
\HERWIG{} preserves as much as possible the four momentum of resonances. A good fraction of the effect can be
tracked back to the final state radiation from the quark, that as remarked previously, is included only by the
shower. Another contribution arises if the hardest jet or the hardest lepton in acceptance are not the ones coming from the leptoquark decay. Of course this happens more easily in showered events.

The plots show that the uncertainty band of the LO predictions vastly underestimates the size of the NLO corrections. There can be considerable shape differences between results at LO and NLO.
The NLO corrections are crucial for an accurate description of these distributions. The $p_T^{\ell j}$ is helpful to discriminate the resonant leptoquark from the single leptoquark plus lepton production~\cite{Dorsner:2018ynv} which features a hard lepton in the production already at tree-level.

\begin{figure}
\centering
{\includegraphics[width=0.475\columnwidth]
   {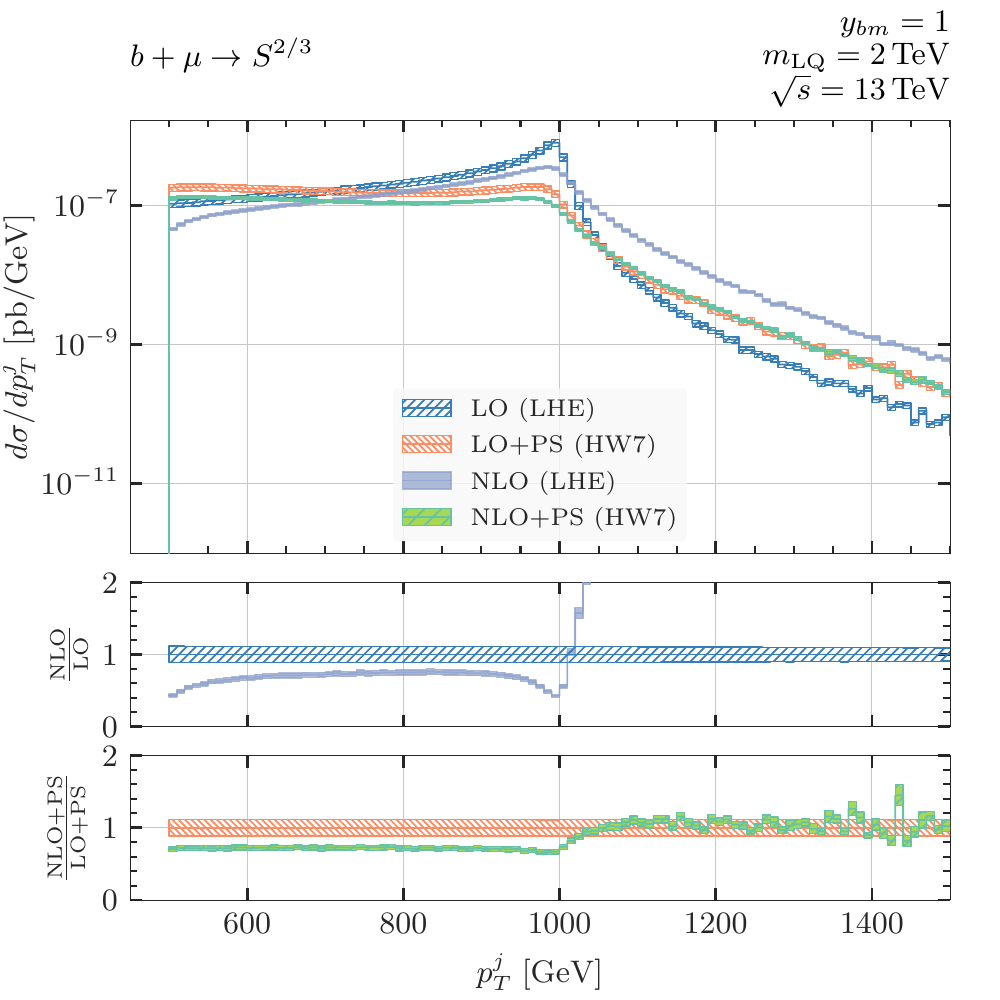}}
{\includegraphics[width=0.475\columnwidth]
   {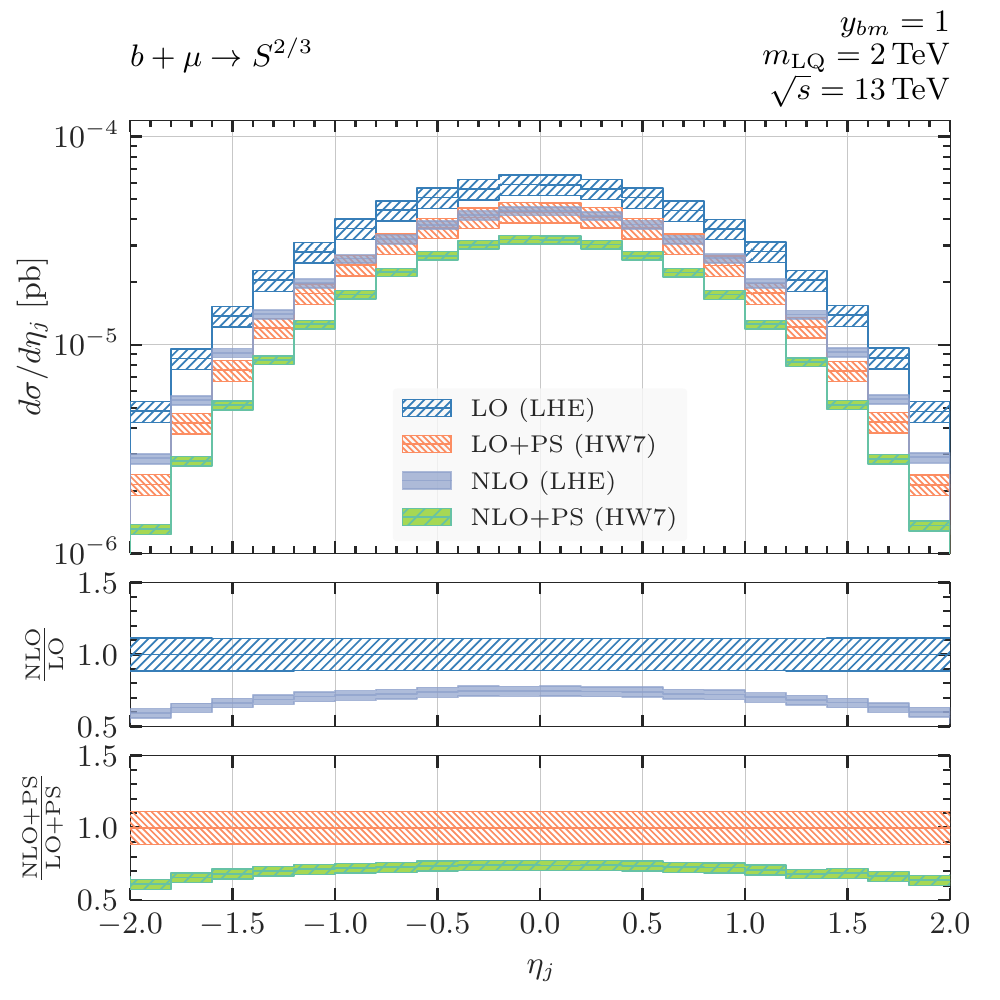}} \\
   \vspace{.1cm}
{\includegraphics[width=0.475\columnwidth]
   {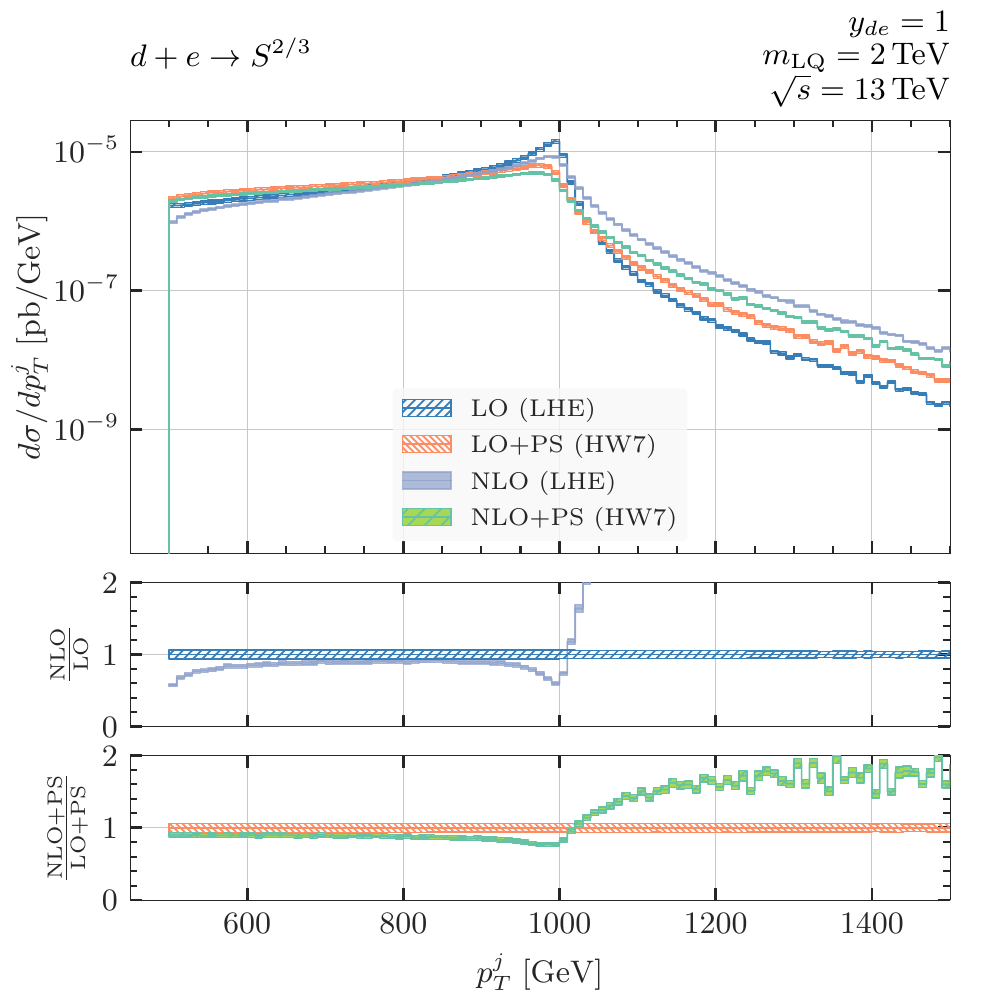}}
{\includegraphics[width=0.475\columnwidth]
   {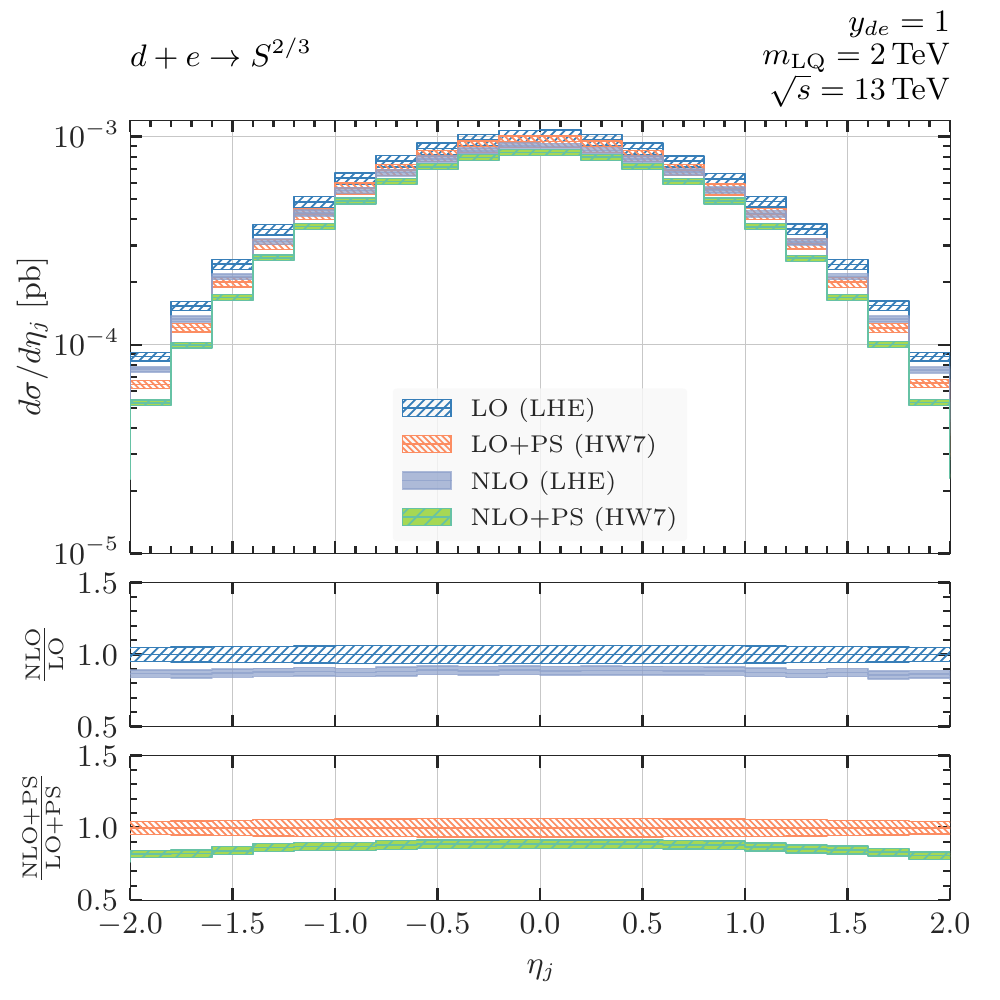}} 
\caption{{\label{fig:plotsjet}} The transverse momentum (left column) and the pseudorapidity (right column) distributions of the leading-$p_T$ jet. See Section~\ref{sec:diffdis} for details.}
\end{figure}

\begin{figure}
\centering
{\includegraphics[width=0.475\columnwidth]
   {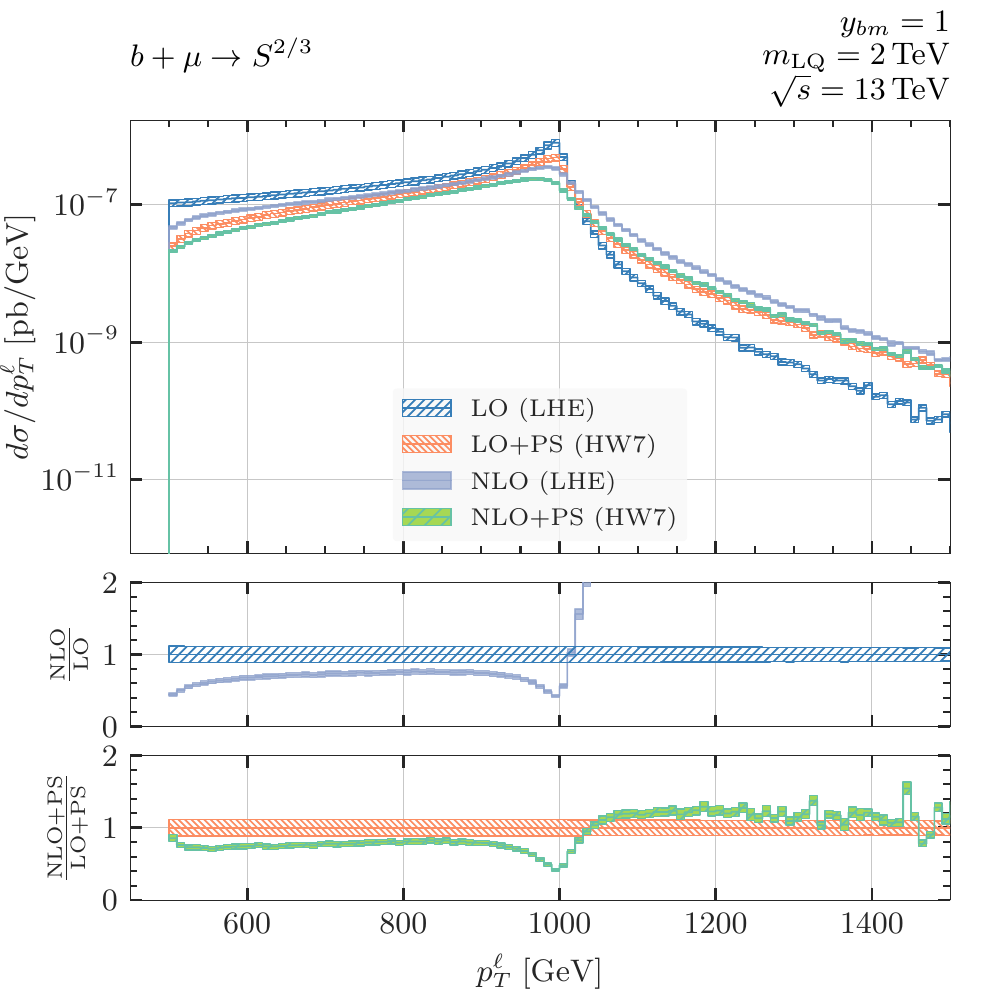}}
{\includegraphics[width=0.475\columnwidth]
   {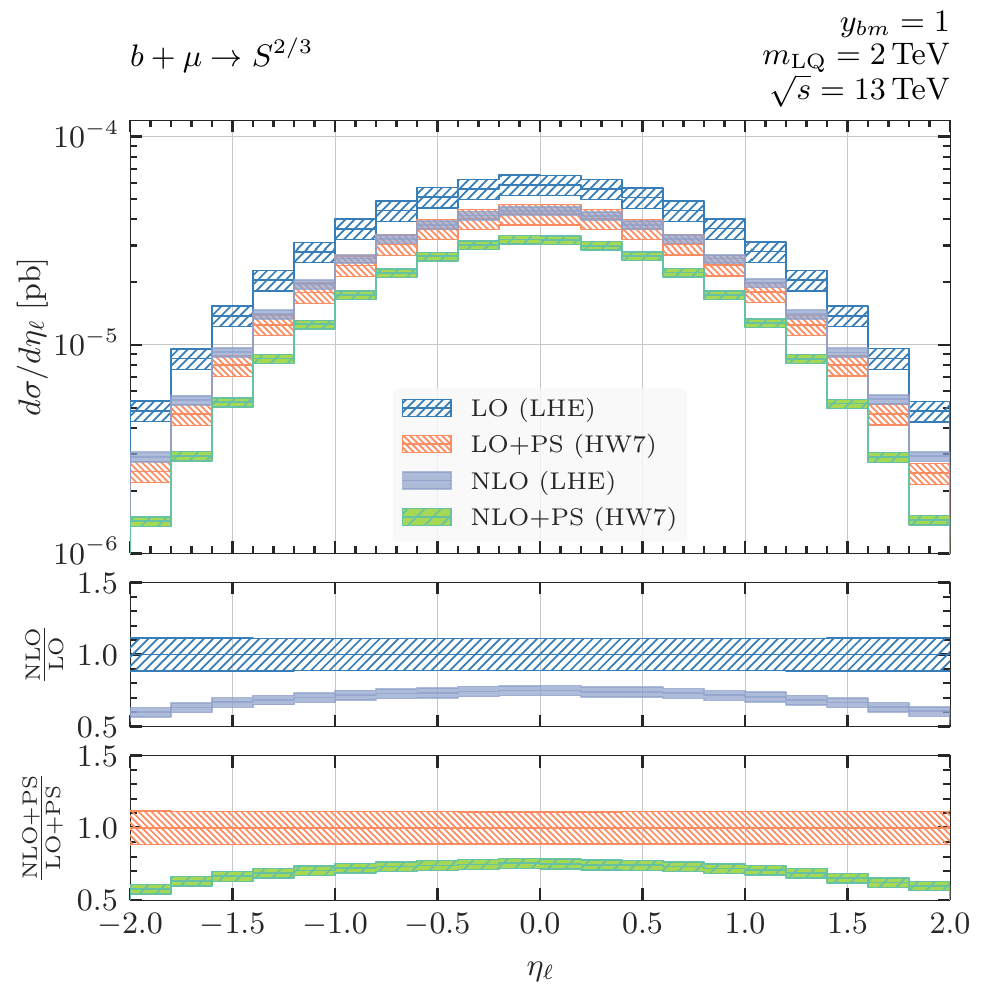}} \\
   \vspace{.1cm}
{\includegraphics[width=0.475\columnwidth]
   {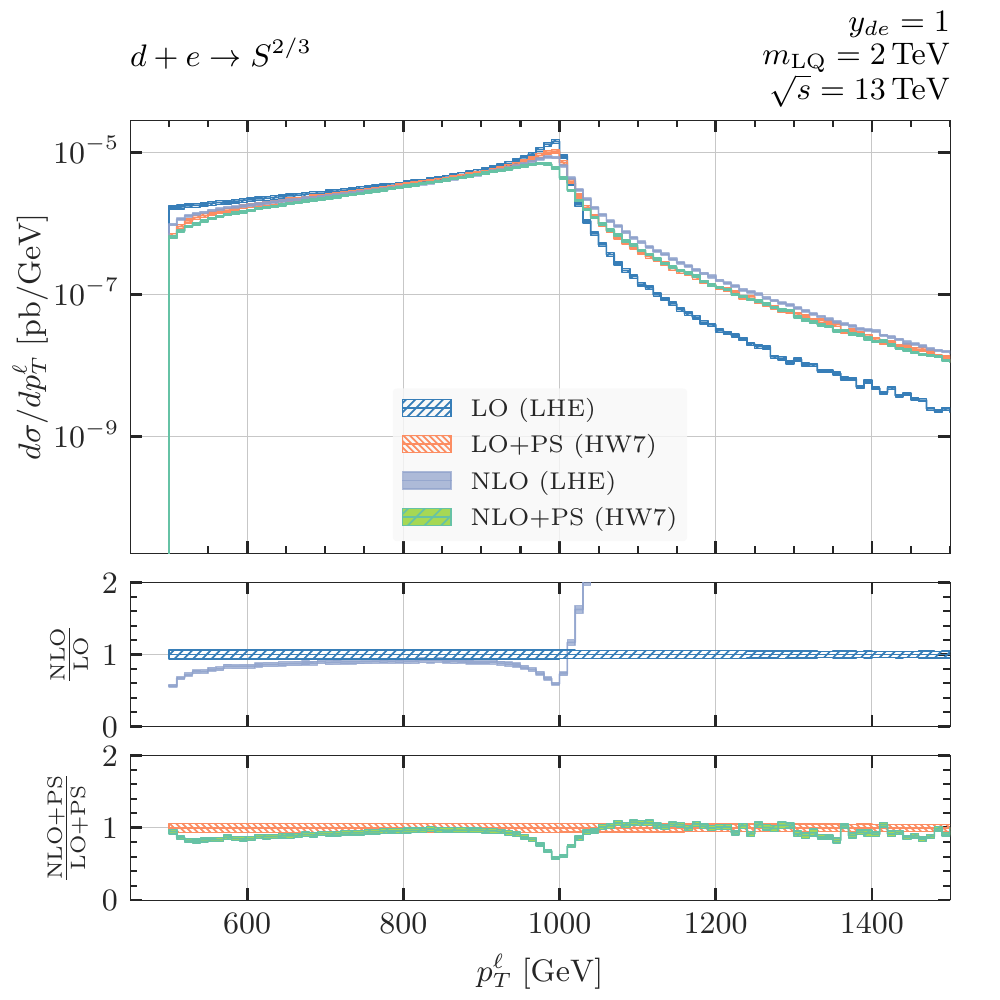}}
{\includegraphics[width=0.475\columnwidth]
   {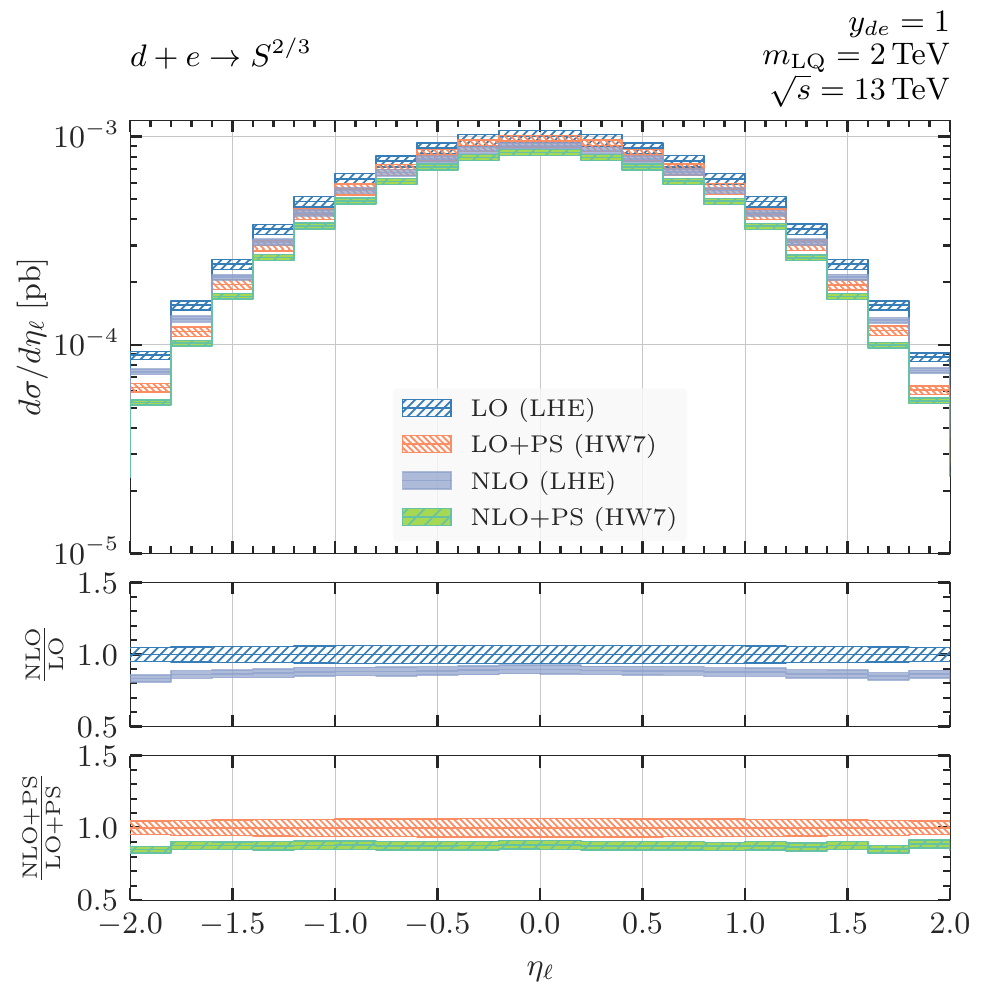}} 
\caption{{\label{fig:plotslep}} The transverse momentum (left column) and the pseudorapidity (right column) distributions of the leading-$p_T$ lepton. See Section~\ref{sec:diffdis} for details.}
\end{figure}

The transverse momentum and the pseudorapidity distributions of the leading jet and lepton are shown in Figure~\ref{fig:plotsjet} and Figure~\ref{fig:plotslep}.
In all plots, higher-order corrections to the pseudorapidity distributions
are slightly flatter than the corresponding ones to the rapidity
distribution of the lepton-jet system. Otherwise, they display a similar
pattern, and similar comments are in place. As expected, the $p_T$ distributions
show a jacobian peak at $m_{{\rm LQ}}/2$. The region above
the kinematic limit $p_T = m_{{\rm LQ}}/2$, strictly forbidden at LO
in the NWA, is populated by finite-width effects. Starting from NLO,
this region also becomes accessible because of extra radiation. This
explains the fact that LO (LHE) predictions for $p_T > m_{{\rm LQ}}/2$
are much softer than the other three predictions. Notice that the NLO (LHE) results' smooth behavior is
generated according to the POWHEG Sudakov factor. Showered
predictions feature a softer spectrum than NLO (LHE) ones due to
final-state radiation. The effect is more pronounced in the case of
the leading jet since the radiation probability for additional QED
emissions is suppressed by the lower coupling $\aem$. This explains
the rise toward smaller $p_T$ values. Comparing the tail of the
leading jet $p_T$ above the peak for the $b+\mu$ and $d+e$ cases, we
observe that, in the latter, NLO+PS(HW7) results present a harder
spectrum than LO+PS(HW7) ones, while they overlap in the former. This
different behavior can be traced back to the interplay between the
initial-state radiation's hardness and a valence quark's presence. 
In fact, in events initiated by a valence quark (that carries a
larger fraction of the proton momentum), the first emission is, on
average harder, and the NLO+PS generator describes this radiation with
higher accuracy than a LO+PS one. On the other hand, in the case of a
sea quark, the same configurations feature a softer initial-state
radiation and a LO+PS description is sufficient to capture the main
effects. For the leading lepton $p_T$ the situation is inverted, with
NLO+PS predictions displaying a harder spectrum than LO+PS ones in the
$b+\mu$ case. The physical mechanism is the same described above with
the difference that, this time, the initial-state lepton colliding
with a sea quark is, on average, more energetic than the one colliding
with a valence quark.

\subsection{Impact on the projected LHC bounds}
\label{sec:sensitivity}

The phenomenological studies performed in~\cite{Buonocore:2020erb,Haisch:2020xjd} disclosed the potential of the resonant leptoquark production through lepton-quark fusion as a competitive search strategy at the LHC, especially for the region of large leptoquark couplings and masses. In that work, the modeling of signal events was based on an approximate LO+PS prediction. The approximation is related to using \PYTHIA{8} to shower the LO events. Indeed, \PYTHIA{8} does not handle lepton-initiated processes. On the other hand, it supports photons in the initial state. Hence, in that work, the particle labels were suitably manipulated to recognize the process as originating from a photon-quark scattering. In this way, the first radiation generated by the shower is likely to be a colored parton most of the time. At the same time, for a lepton-quark scattering event, the photon splitting process $\gamma \to \ell {\bar \ell}$ competes with QCD radiation in the backward evolution.\footnote{The pdf ratio $f_\gamma/f_\ell \sim
\alpha_s/\alpha$ compensates for the factor of $\alpha$ arising from the photon splitting.} The resulting mismodeling of the hadronic activity in the event was estimated to have only a mild impact, affecting the prediction for the reconstructed leptoquark mass by roughly $20\%$~\cite{Buonocore:2020erb}.

In the present work, we have improved the simulation of the signal
events in two ways: first, we include the full set of NLO corrections
to the leptoquark production process, from now on NLO$_{P}$, and, second, we
match \emph{\`a la} POWHEG the NLO$_P$ corrections to a modified version of the
\HERWIG{7} parton shower~\cite{Bellm:2019zci} that handles lepton initiated
processes~\cite{HW7dev}. In the following, we assess the relative impact of these
improvements. We consider as a benchmark point a scalar leptoquark of nominal
mass $m_{{\rm LQ}}=3\,$TeV, charge $Q_{{\rm LQ}}=\pm 1/3$ and which couples only
to electrons and up quarks. 

Our main focus is on the reconstructed jet-lepton invariant mass where the leptoquark shows up as a resonance. Additional selection cuts are
crucial to tame the SM background. However, they largely affect the shape of
the resonant peak and the acceptance of signal events. We consider a simplified version
of the fiducial volume defined in~\cite{Buonocore:2020erb} to analyze the main radiative effects. As a
basic requirement, dubbed as \emph{cut A}, we select events with at
least one lepton and one jet in the central region of the detector,
$|\eta^{\ell,j}|<2.5$. We then impose the following set of cuts on
leading and subleading leptons/jets, collectively referred to as \emph{cut
B}: $p_{T}^{\ell_1,j_1}>500\,$GeV, a veto on secondary leptons with
$p_T^{\ell_2}>7\,$GeV and $|\eta^{\ell_2}|<2.5\,$, a veto on secondary
jets with $p_T^{j_2}>30\,$GeV and $|\eta^{j_2}|<2.5\,$.

\begin{figure}[!ht]

\includegraphics[width=0.45\columnwidth]{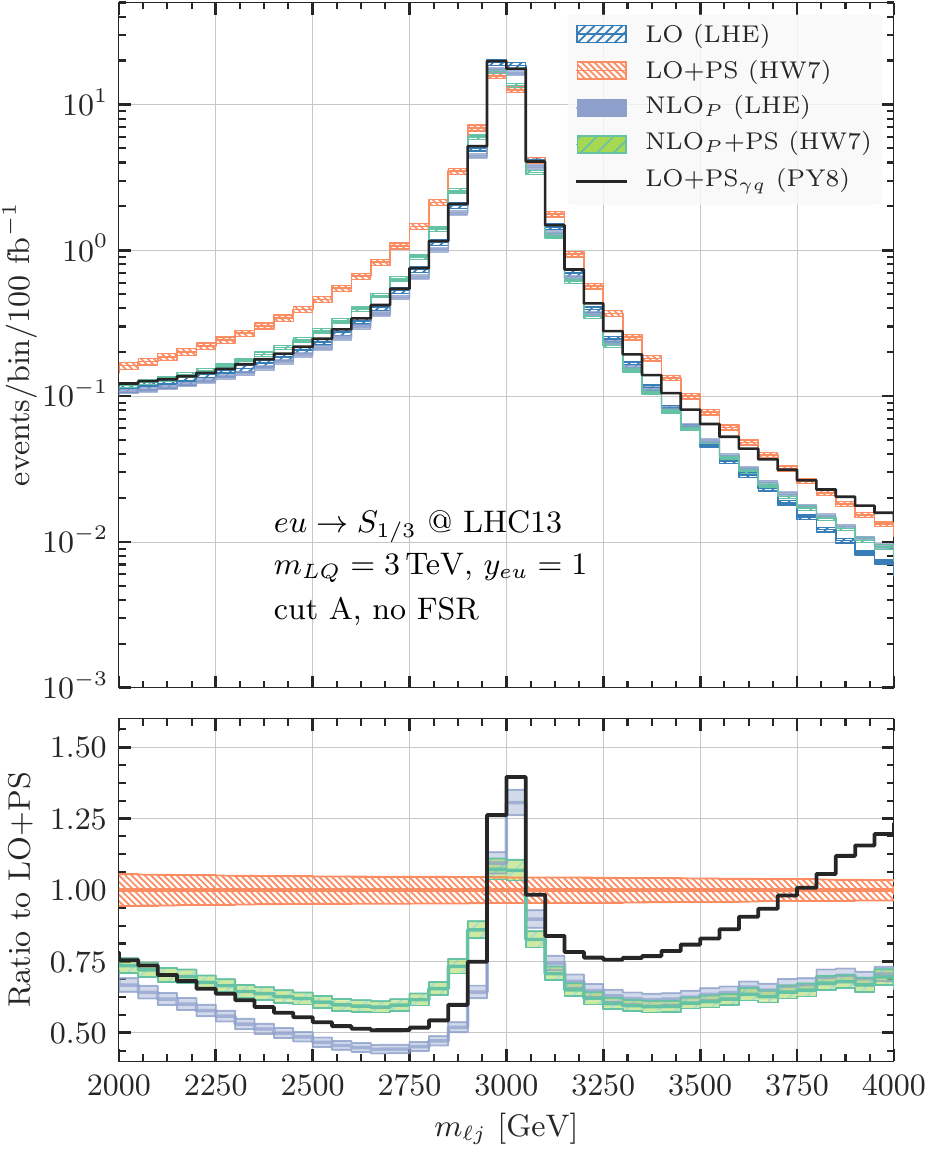}
\hfill
\includegraphics[width=0.45\columnwidth]{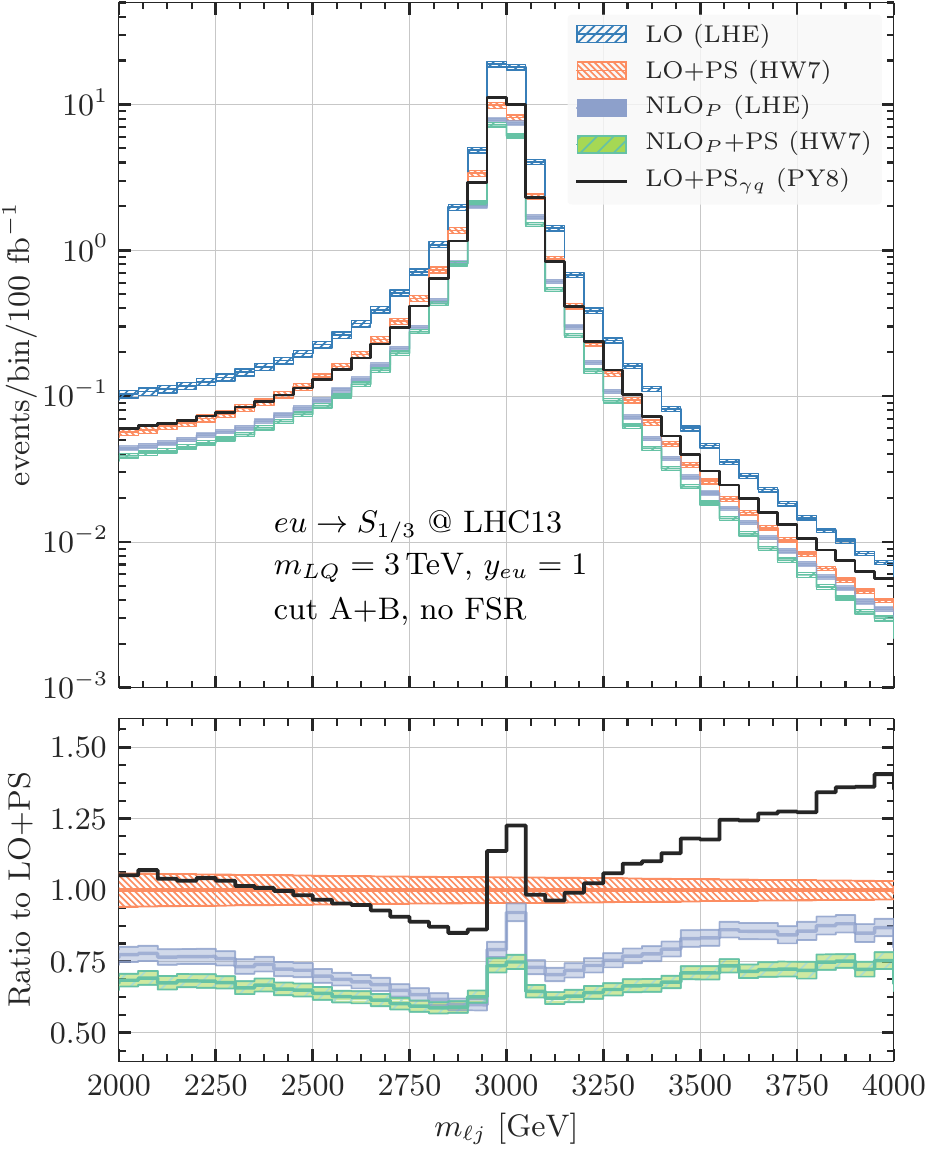}
  
\vspace{0.5cm}

\centering
\includegraphics[width=0.45\columnwidth]{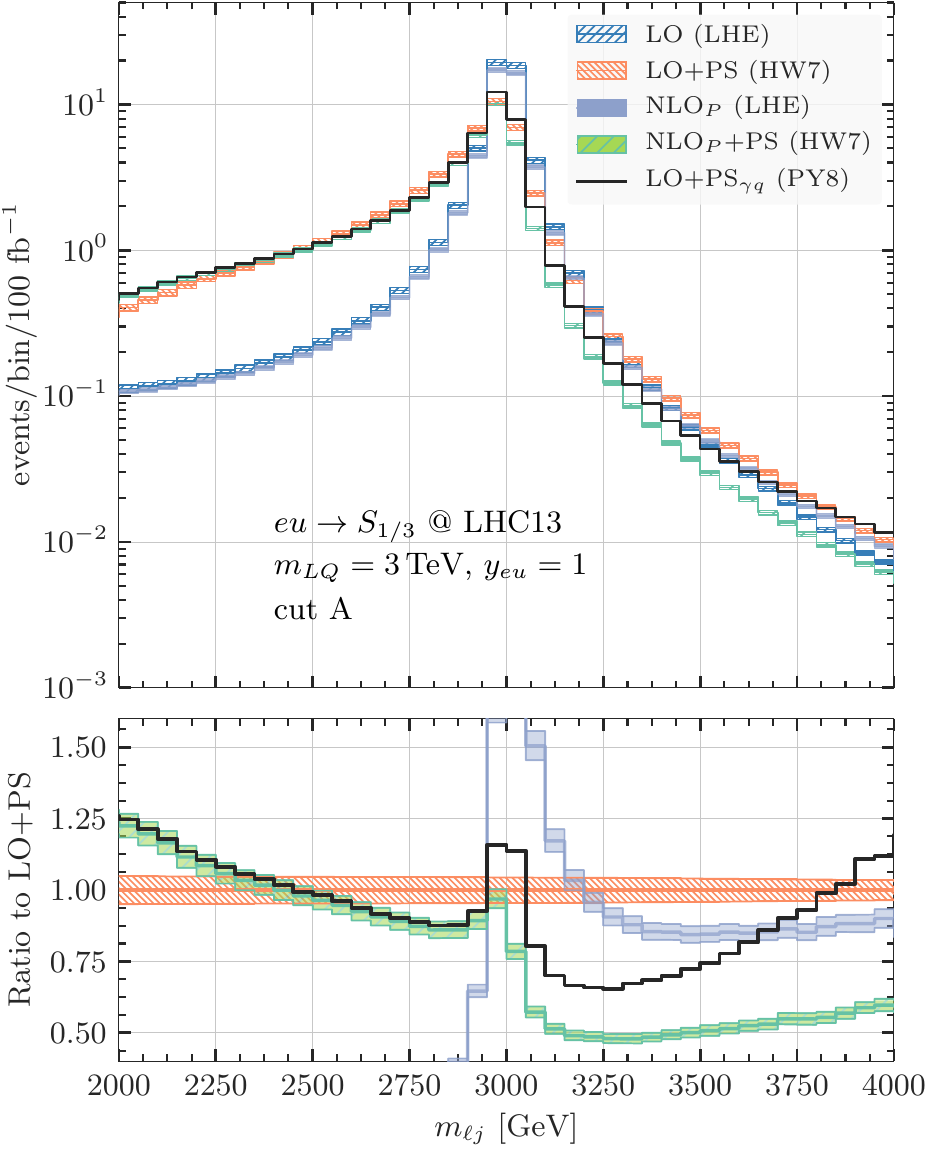}
\hfill
\includegraphics[width=0.45\columnwidth]{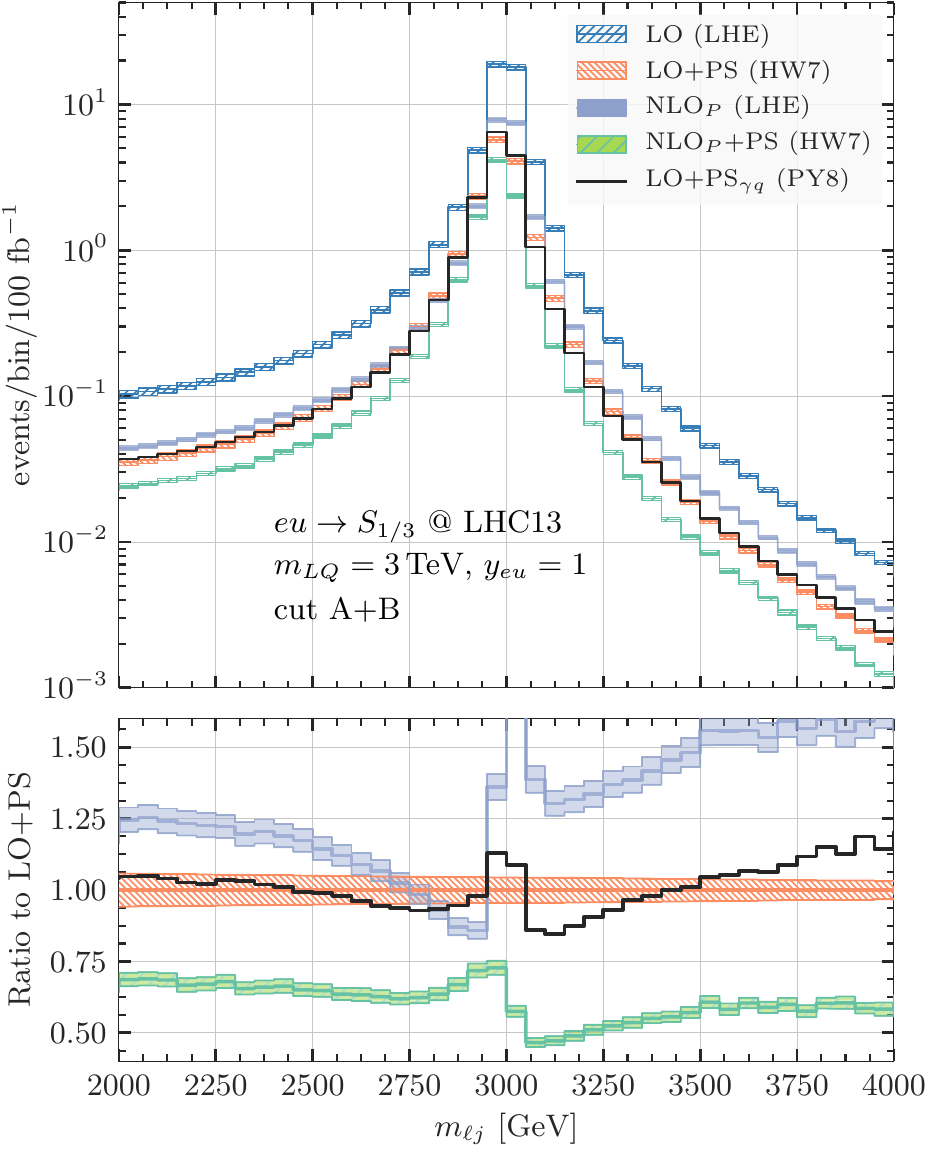}
\vspace{0.2cm}
\caption{ \label{fig:recmass} Invariant mass distribution of the leading-$p_T$ lepton and jet system, $m_{\ell j}$, in a model of resonant $s$-channel leptoquark production through lepton-quark fusion at the LHC. In each figure, the various predictions correspond to generators at different accuracy, as explained in the main text, while the bottom panel contains the ratio to the LO+PS~(HW7) one. We display results for the number of events requiring only \emph{cut A} (left) and the combination \emph{cut A+B} (right). Radiation from the decay products is disabled in the two top figures.}
\end{figure}

In Figure~\ref{fig:recmass}, we compare different predictions for the
invariant mass distribution of the system composed of the hardest
lepton and hardest jet, obtained with samples of signal leptoquark events at
different accuracy: LO (blue), LO+PS~(HW7) showered with \HERWIG{7}
(orange), NLO$_{P}$ Les Houches events as generated with \POWHEG{}
(gray), NLO$_{P}$+PS~(HW7) the same events showered with \HERWIG{7} (green). The bands
correspond to the customary 7-point scale variation. We also report in
black the LO+PS$_{\gamma q}$~(PY8) prediction obtained showering the
events with \PYTHIA{8} after performing the replacement of
the initial lepton with a photon as done in~\cite{Buonocore:2020erb}. We leave out multiparton interactions (MPIs) and detector effects from
these comparisons to facilitate the discussion. Furthermore, we do not apply any recombination of
photons with a close-by lepton.

Let us remind the reader that we computed NLO radiative corrections only to the leptoquark production process, leaving to the parton shower the full description of the radiation from the decay products. For this reason, it is interesting to consider first the case in which we
switch off final-state radiation (FSR) in the parton shower, which more
closely resembles the radiative content of our NLO$_{P}$ prediction. We
start focusing on the plots of the left-hand side of
Figure~\ref{fig:recmass}, where we apply only the essential requirement
\emph{cut A}. Comparing top and bottom, we observe that the
exclusion of FSR leads to much milder parton shower effects. Indeed,
the distinctive radiative tail in the bottom plot is entirely due to
QCD FSR, which forms a separate second jet softening the leading jet,
originated by the quark in the leptoquark decay. Furthermore, we have
explicitly verified that photon-to-lepton recombination has a minimal impact on the distribution, confirming that the FSR effects due
to QED radiation are less important. When FSR is not included, all
predictions are close-by among each other within $15-20\%$, except for the one obtained with LO+PS~(HW7) generator. This
might be due to different shower mechanisms and recoil prescriptions
in \PYTHIA{8} and \HERWIG{7}, whose impact becomes less prominent after
performing the matching to the NLO$_{P}$ computation. While this is an
interesting topic, its investigation is beyond the aim of the present
work, and it is left for a future study.

\begin{table}
  \centering
  \begin{tabular}{ccccc}
    \toprule
      & \emph{cut A} (noFSR) & \emph{cut A+B} (noFSR) & \emph{cut A} & \emph{cut A+B} \\
    \midrule
    LO & 0.96 & 0.89 & 0.96 & 0.89  \\ 
    LO+PS (HW7) & 0.98 & 0.48 & 0.98 & 0.28 \\
    NLO$_{P}$ & 0.97 & 0.42 & 0.97 & 0.42 \\
    NLO$_{P}$+PS (HW7) & 0.98 & 0.37 & 0.99 & 0.20  \\
    LO+PS$_{\gamma q}$ (PY8) & 0.97 & 0.51 & 0.98 & 0.29 \\
    \bottomrule
  \end{tabular}
  \caption{ \label{tab:acceptance} The cut flow analysis. The table shows the acceptance $A=\sigma_{\rm cut}/\sigma_{\rm no cuts}$ associated to cut A and cut A+B. See Section~\ref{sec:sensitivity} for details.}    
\end{table}

We turn to the more interesting situation in which we apply the
combination of cuts \emph{cut A+B}. The results are shown in the plots
of the right-hand side of Figure~\ref{fig:recmass}, excluding (top) or
not (bottom) FSR radiation. Since the cuts are tailored to enhance
Born-like configurations, the LO prediction remains in practice
untouched. On the other hand, the veto on secondary leptons and jets
vastly reduces all the other predictions, see Tab~\ref{tab:acceptance}. In addition, the radiative tail seen in
the more inclusive setup is effectively cut out. In particular, when
FSR is excluded (top right), we observe that the result obtained with
the NLO$_{P}$+PS~(HW7) generator only mildly differs from the
NLO$_{P}$ one, meaning that only the first few emissions are relevant
for the computation of the acceptance. The radiation from the decay
products further reduces the acceptance. Since the NLO$_{P}$
computation does not contain such effects, it fails to describe the
total result. Instead, the LO+PS predictions and the more accurate
NLO$_{P}$+PS one includes radiative effects from the decay as modeled
by FSR of the parton shower (bottom right).

The main results regarding the A+B cuts can be summarised as follows:
\begin{itemize}
\item NLO$_{P}$ provides an estimate of the acceptance that, however, 
  misses the effects due to radiation from the decay products;
\item LO+PS$_{\gamma q}$~(PY8) and LO+PS~(HW7) give results in
  reasonable agreement among each other, with very mild differences of
  about $15\%$. Nonetheless,
    we notice that this result might be accidental given that we
    observe substantial differences in the mass spectrum predicted by
    the two generators when FSR is turning off. This issue seems
    alleviated after the NLO matching, as the NLO$_P$+PS(HW7) and
    LO+PS$_{\gamma q}$~(PY8 display an overall better agreement in
    shape. We believe that this is a further motivation for the
    experts in parton showers to pursue the study of lepton initiated
    processes in proton-proton collisions;
 \item by comparing with the NLO$_P$, FSR radiation contributes to the reduction of the
  acceptance of a further $50\%$, see also Tab.~\ref{tab:acceptance};
\item the most accurate NLO$_{P}$+PS prediction leads to a further reduction
  of the acceptance of about $30\%$ with respect to LO+PS ones. This
  can be explained by the fact that the former includes the exact matrix
  element for the first emission in production. As a result, we expect the limits on the leptoquark couplings shown in Figure 3 of~\cite{Buonocore:2020erb} to relax by about $15\%$.
\end{itemize}

It is well known that parton showers usually provide a better
description of FSR radiation than ISR. Therefore, one may expect that the
NLO$_{P}$+PS description computed in this work already captures the
main radiative effects. Nonetheless, given the importance of FSR in
computing the acceptance, a natural extension of the present work
would be to match the NLO computation for all possible resonant and
non-resonant $2\to2$ quark-lepton processes to the parton shower,
thus including radiation from all legs and (or) resonant intermediate
states.

\subsection{The case study: $S_3$ leptoquark}
\label{sec:S3example}

To illustrate the usage of the code for a particular UV model, we add to the SM an additional scalar field transforming in the anti-fundamental of $SU(3)_{{\rm C}}$ and the adjoint of  $SU(2)_{{\rm L}}$ with the hypercharge $Y=1/3$, known as the $S_3\sim(\bar{\textbf{3}},\textbf{3})_{1/3}$ leptoquark~\cite{Dorsner:2016wpm}. In the unbroken phase, the renormalisable Lagrangian describing the couplings of $S_3$ to the SM fermions reads
\begin{equation}
-\mathcal{L} \supset \lambda_{q\ell} \,{\bar{Q}}_L^{Ca} \epsilon^{ab} (\sigma^k S_3^k)^{bc} L_{L}^{c} + \lambda_{qq}\,{\bar{Q}}_{L}^{Ca}  \epsilon^{ab}((\sigma^k S_3^k)^\dagger)^{bc} Q_{L}^{c} + \rm{h.c.}\,,
\end{equation}
where $\sigma^{k=1,2,3}$ are the Pauli matrices, $\epsilon^{ab}=(i \sigma^2)^{ab}$,  $C$ stands for charge conjugation, and $S_3^k$ are leptoquark components in the $SU(2)_{{\rm L}}$ space. The matrices $\lambda_{q\ell}$ and $\lambda_{qq}$ are generic $3\times3$ matrices in flavour space. The summation over flavours is assumed. It is easy to argue that dangerous diquark couplings $\lambda_{qq}$ are absent due to an (approximate) baryon number conservation.\footnote{This can be achieved, for example, in some GUT models where $S_3$ is embedded in $SO(10)$ or $SU(5)$ irreducible representation~\cite{Senjanovic:1982ex,Dorsner:2017ufx}. Another example is to gauge a lepton flavour non-universal $U(1)$ under which $S_3$ is charged such that $\Delta B = 0~({\rm mod}~3)$  completely forbids proton decay~\cite{Davighi:2022qgb}.}

The left-handed  quark and lepton $SU(2)_{\rm L}$ doublets, $Q_L$ and $L_L$, are assumed to be in the down-quark and charged-lepton mass basis, respectively. After the electroweak symmetry breaking, the relevant interactions of the electromagnetic charge eigenstates $S_{-2/3}=(S_3^1+iS_3^2)/\sqrt{2}$, $S_{1/3}=S_3^3$, and $S_{4/3}=(S_3^1-iS_3^2)/\sqrt{2}$, in the notation of Eq.~\eqref{eq:Lagdef}, read
\begin{align}
\label{eq:S3_broken}
-\mathcal{L}&\supset y^L_{U\nu}\, \bar{U}^C_L \nu_L\, S_{-2/3}+y^L_{U\ell}\, \bar{U}^C_L  \ell_L\, S_{1/3}+y^L_{D\nu}\, \bar{D}^C_L \nu_L\, S_{1/3} 	+y^L_{D\ell}\, \bar{D}^C_L  \ell_L\, S_{4/3} + \rm{h.c.}\,,
\end{align}
where $U$ and $D$ stand for the three up- and down-type quarks, while $\ell$ ($\nu$) stands for the three charged leptons (neutrinos), and
\begin{align}
y^L_{U\nu} &= \sqrt{2} V^T_{{\rm CKM}} \lambda_{q\ell}
	V_{{\rm PMNS}}
	\,,\quad
	y^L_{U \ell} = - V^T_{{\rm CKM}} \lambda_{q\ell}
	\,,\\
	y^L_{D\nu} &= \lambda_{q\ell} V_{{\rm PMNS}}
	\,,\quad\quad\quad\quad\quad
	y^L_{D \ell} =- \sqrt{2} \lambda_{q\ell}
	\,,
\end{align}
where $V_{{\rm PMNS}}$ is the Pontecorvo–Maki–Nakagawa–Sakata mixing matrix, and $V_{{\rm CKM}}$ is the Cabibbo-Kobayashi-Maskawa mixing matrix.  The $SU(2)_{{\rm L}}$ gauge symmetry predicts the three states to be nearly mass-degenerate. Potentially significant contributions to the mass splitting are constrained by the electroweak precision tests~\cite{Dorsner:2016wpm}. This is, of course, very important for the direct searches at the LHC, predicting multiple degenerate resonances.

Since the neutrino PDF in the proton is vanishing (at the order in perturbation theory we are working at\footnote{It can be generated with a mechanism similar to the lepton PDF, going through a $Z/W$ rather than a photon, but, unlike the lepton case, the logarithmic enhancement is missing.}), proton collisions can produce only the states with charges $1/3$ and $4/3$ in the quark-lepton fusion. However, various decay channels are generally open (including neutrinos), and the branching ratios depend on the flavour structure of $\lambda_{q\ell}$. 

When the leptoquark flavour matrix $\lambda_{q\ell}$ has an anarchic structure, the low-energy flavour physics observables set a lower limit on $m_{{\rm LQ}}$ to be far above the TeV scale, see~\cite{Dorsner:2016wpm,Isidori:2010kg}. A consistent scenario should therefore exhibit flavour protection. For simplicity, we assume that the leptoquark carries a global $U(1)_{j} \times U(1)_{\beta}$ quark and lepton charges, where $j$ and $\beta$ denote a particular quark and lepton flavour combination, such that the only allowed coupling becomes
\begin{equation}
  \lambda_{q\ell}  \xrightarrow{U(1)_j \times U(1)_{\beta}}\lambda\, \delta_{q j} \delta_{\ell\beta}\,,
\end{equation}
where $\lambda$ is a complex number. For example, the case in which the leptoquark is charged under the global $U(1)_{1} \times U(1)_{2}$ symmetry implies that the only non-vanishing entry in $\lambda_{q\ell}$ is the $1-2$ entry, $\lambda_{12} = \lambda$. Neglecting neutrino masses, which is an excellent approximation at relevant energies, this symmetry is broken only by the CKM mixing matrices. In this limit, the flavour-changing contributions in the quark sector are suppressed by the smallness of the off-diagonal CKM elements while charged lepton flavour is exactly conserved. For the direct searches at the LHC, the CKM can safely be approximated with the unit matrix. 

\begin{figure}[!t]
	\centering
	\includegraphics[scale=0.35]{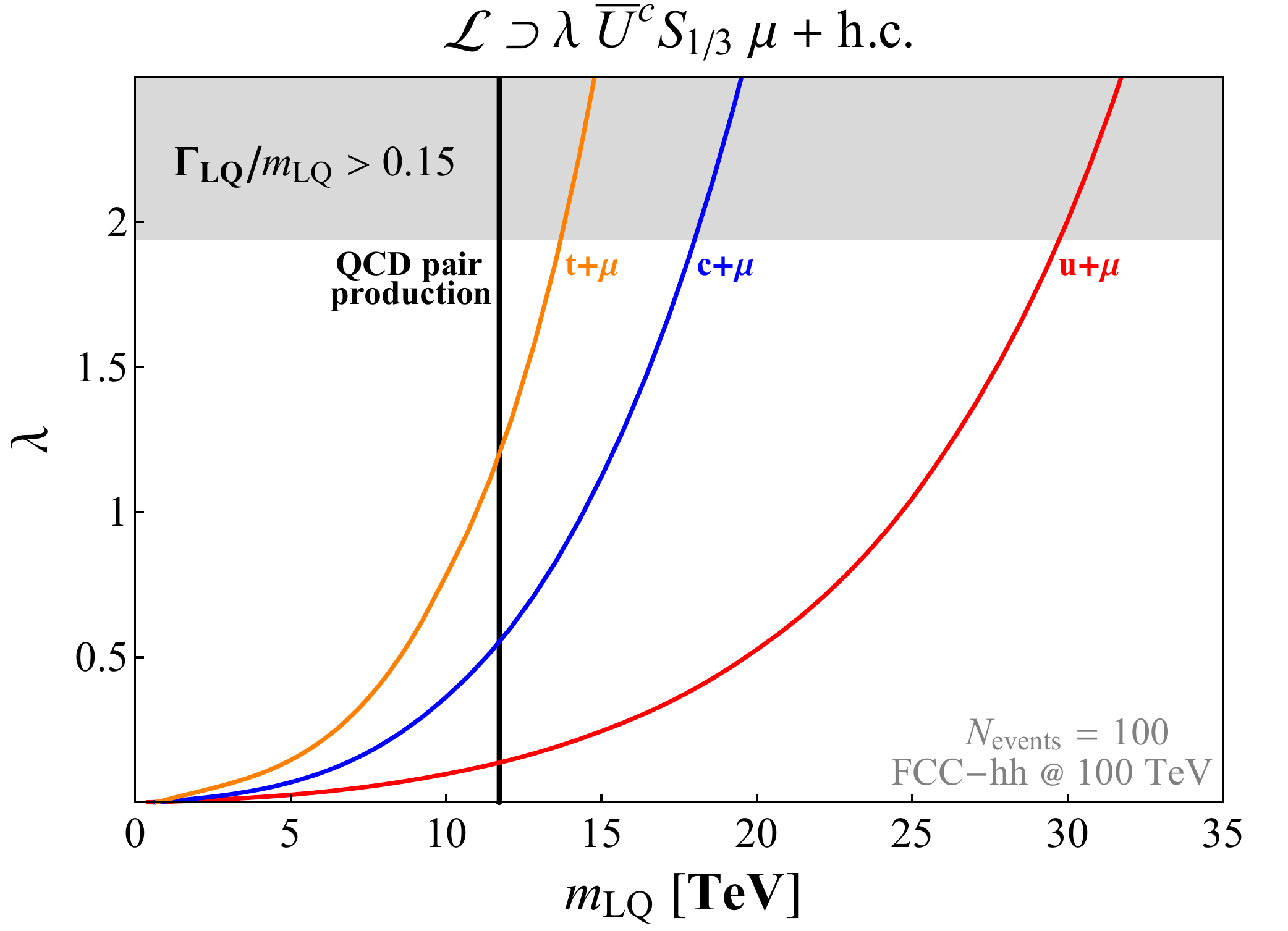}
	\includegraphics[scale=0.35]{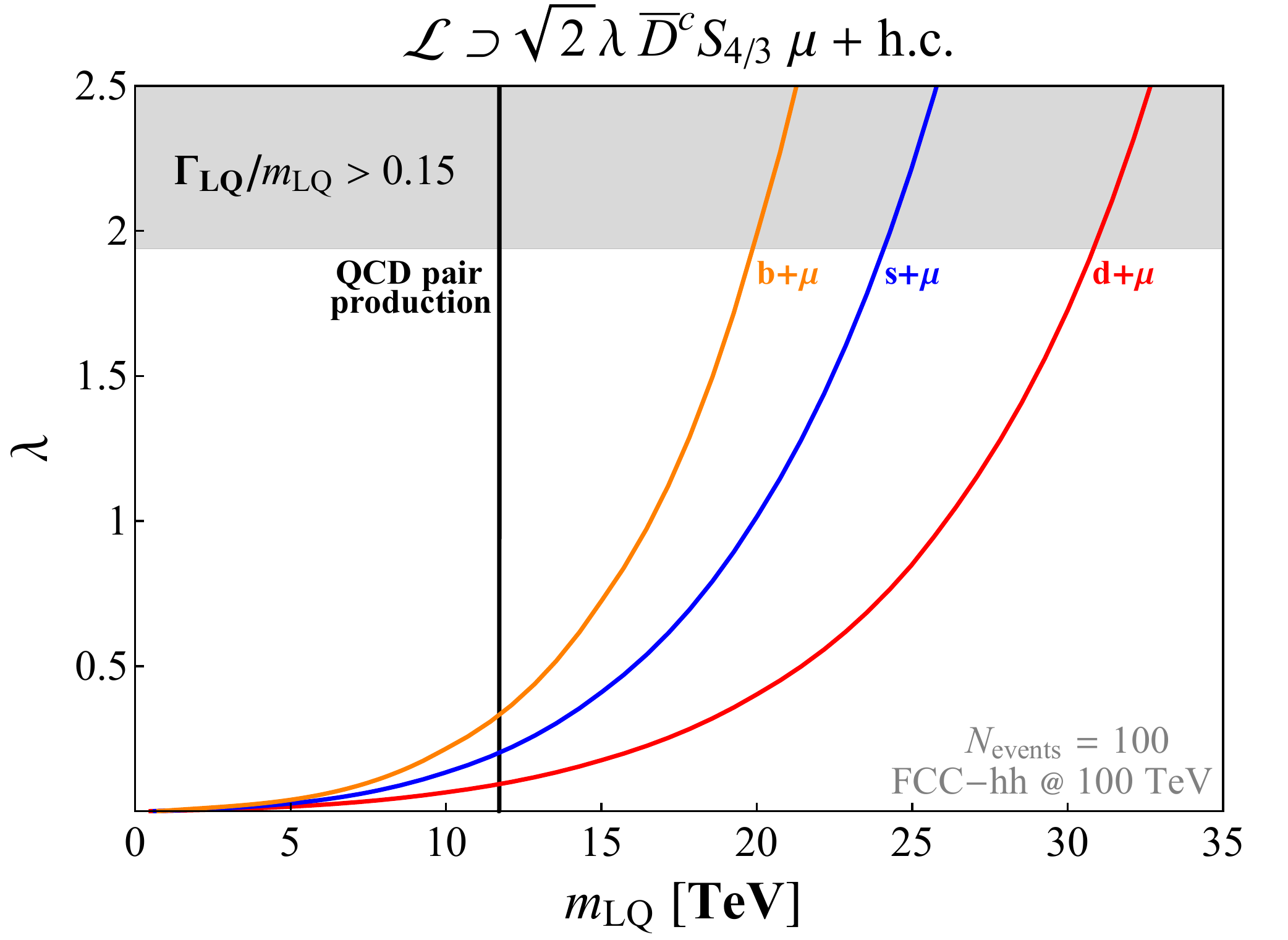}
	\caption{Contours in the $(m_{{\rm LQ}},{\lambda})$ plane for producing 100 events at the FCC-hh for $\pm1/3$ (left plot) and $\pm4/3$ (right plot) charge components of the $S_3$ scalar leptoquark model from Section~\ref{sec:S3example}. The solid black line is for the QCD pair production, while the red, blue, and orange are for the resonant leptoquark production, assuming couplings to the first, second, or third generation of quarks, respectively. The grey shaded region shows the regime of a broad resonance.}
	\label{fig:FCChh}
\end{figure}

The bottomline of these assumptions is that the leptoquark interacts dominantly with a single generation of quarks and a single generation of leptons. In the following, we will study all six quark flavour cases separately. Since the lepton PDF are similar across different flavours, we will consider only the coupling to muons for simplicity.\footnote{An example of a particularly motivated flavour structure in the quark sector is $U(2)^3_q$ flavour symmetry under which the third generation is invariant while the light generations form doublets~\cite{Barbieri:2011ci,Faroughy:2020ina,Greljo:2022cah}. This symmetry is an excellent approximate symmetry of the SM Yukawa sector. In the leptonic sector, the $U(1)_{\mu-\rm{LQ}}$ symmetry can result accidentally from a lepton non-universal gauge symmetry~\cite{Greljo:2021xmg,Greljo:2021npi,Davighi:2020qqa,Hambye:2017qix,Davighi:2022qgb,Heeck:2022znj}. Thus, in the exact symmetry limit, only the $3\,-\,2$ entry, $\lambda_{32}$, is allowed. Figure~\ref{fig:FCChh} shows this case with the orange curve, while the projections at future colliders (in other channels) were also considered in~\cite{Azatov:2022itm}.}

With all this, the LO decay widths of $S_{1/3}$ and $S_{4/3}$ states are given as
\begin{align}
\label{eq:width1/3}
\Gamma_{S_{1/3}}  & =\frac{(y_{U \ell}^L)^* y_{U \ell}^L+(y_{D\nu}^L)^* y_{D\nu}^L}{16\pi}\,m_{\rm{LQ}} \quad \xrightarrow{U(1)_{j} \times U(1)_{2}} \quad \frac{|\lambda|^2}{8\pi}\,m_{\rm{LQ}}\,,\\
\label{eq:width4/3}
\Gamma_{S_{4/3}} &  =\frac{(y_{D \ell}^L)^* y_{D \ell}^L}{16\pi}\,m_{\rm{LQ}} \quad \xrightarrow{U(1)_{j} \times U(1)_{2}} \quad \frac{|\lambda|^2}{8\pi}\,m_{\rm{LQ}}\,,
\end{align}
where we sum over quark and lepton flavour indices. In the case of $U(1)_{j} \times U(1)_{2}$ global symmetry,  $\lambda = \lambda_{j2}$, with $j=1,2,3$ depending on the quark generation which couples to the leptoquark.

Let us finally discuss the importance of the resonant leptoquark production at the FCC-hh collider operating at $100\,\rm{TeV}$ proton-proton center of mass energy, with the luminosity of $30\,\rm{ab}^{-1}$. We do not aim to derive precise projections since a complete analysis, including the signal and background simulations, is clearly beyond the scope of this work. Instead, a simple comparison with the QCD pair production can already be made using the inclusive cross sections from Section~\ref{sec:xsec} and predicted branching ratios to determine the parameter space for which one can produce more than 100 events. Even though it is a naive estimate, since the SM background is subleading in the high-energy bins (as proved in~\cite{Buonocore:2020erb}) and the signal is resonant, we expect 100 events to guarantee a discovery. A detailed projection study for the QCD pair production mechanism~\cite{Allanach:2019zfr} gives a result very close to this criteria.

Figure~\ref{fig:FCChh} illustrates the main point. The plot on the left (right) side is for the $\pm1/3$ ($\pm4/3$) state. The solid black line is for the QCD pair production, while the red, blue, and orange lines are for different quark generations in increasing order. The regions left to the lines is where the FCC-hh can produce more than 100 events. Finally, the grey shaded region predicts a broad resonance ($\Gamma_{{\rm LQ}} / m_{{\rm LQ}} > 0.15$). The plots show large portions of parameter space for which the resonant leptoquark production channel offers a unique window for discovery. This finding motivates a comprehensive projection study for future work.

\section{Conclusions}
\label{sec:conclu}

Leptoquarks at the TeV scale are predicted in various settings beyond the SM, such as non-minimal composite Higgs models, $R$-parity violating supersymmetric models, extended gauge symmetries, and others. Leptoquark extensions of the SM have recently been under the spotlight as promising candidates to address various flavour anomalies. As a result, ATLAS and CMS collaborations are investing increasingly more resources to search for these particles. Direct discovery of a leptoquark would have profound implications for the paradigm of quark-lepton unification at shorter distances.

Precise determination of lepton densities inside the proton~\cite{Buonocore:2020nai} revealed a novel path for leptoquark production at the LHC. Despite the smallness of the lepton PDF, a direct quark-lepton fusion at the partonic level is the most sensitive production channel for large leptoquark couplings, thanks to the resonant enhancement. Indeed, the first phenomenological studies show that large portions of the leptoquark model's parameter space can uniquely be probed through this channel~\cite{Buonocore:2020erb}. This study, however, relies on tree-level calculations and a crude estimate of the lepton shower effects that were not developed at the time. The first calculation at NLO~\cite{Greljo:2020tgv}, albeit limited to the inclusive cross sections, showed an interesting pattern of QCD and QED corrections which are similar in size. However, the full NLO description of the process, including differential distributions, was still missing.

In this work, we develop the first Monte Carlo event generator for precision studies of the resonant leptoquark production at hadron colliders. In Section~\ref{sec:setup} we present the \POWHEGBOXRES{} implementation of the process at NLO matched to parton shower, including the lepton shower, which has recently become available in \HERWIG{}. Section~\ref{sec:decay} discusses leptoquark decays and the treatment of the resonance line shape. Our code allows for a full-fledged simulation of the process and is flexible enough to include all renormalisable scalar leptoquark models with arbitrary flavour structures~\cite{Dorsner:2016wpm}. We leave for future work the implementation of the vector leptoquark models. 

We validate the code by reproducing the NLO inclusive cross sections at the LHC~\cite{Greljo:2020tgv} and provide new results for the FCC-hh (see Section~\ref{sec:xsec}). The unique advantage of our \POWHEGBOXRES{} implementation is the possibility to study arbitrary differential distributions. In Section~\ref{sec:diffdis}, we comprehensively investigate the phenomenologically relevant observables, such as the jet-lepton azimuthal distance and the system's invariant mass, $p_T$, and rapidity. As an illustration, in Figures~\ref{fig:mLQyLQ} and \ref{fig:dphiptlq} we show these distributions for three different benchmark points, at LO or NLO and with or without PS. We conclude that higher-order corrections are kinematics-dependent and should be adequately incorporated. In Section~\ref{sec:sensitivity}, we study the importance of improved signal predictions on the (HL-)LHC projections reported in~\cite{Buonocore:2020erb}. We closely follow the cut flow analysis of~\cite{Buonocore:2020erb} to find an overestimation of the acceptance of up to $30\%$.

 To illustrate the potential of the FCC-hh, in Section~\ref{sec:S3example} we study a concrete model, the $S_3\sim(\bar{\textbf{3}},\textbf{3})_{1/3}$ scalar leptoquark. Figure~\ref{fig:FCChh} shows the leptoquark coupling as a function of mass needed to produce 100 events. We find that the resonant production mechanism can potentially probe uncharted parameter space beyond the reach of QCD pair production for all quark flavours (including the top quark). Our simplified analysis motivates a detailed projections study at the FCC-hh, including the background simulation, which is left for future work.

To conclude, this work paves the way for the first experimental searches and further phenomenological studies of the resonant leptoquark production at the LHC (and beyond). The latest addition to the leptoquark toolbox is made publicly available at the website \url{http://powhegbox.mib.infn.it}.

\section*{Acknowledgements}

We thank Silvia Ferrario Ravasio for clarifying issues with \HERWIG. The code for the computation
of the running $\aem$ in the project repository was taken from the Hoppet code~\cite{Salam:2008qg}. The work of AG has received funding from the Swiss National Science Foundation (SNF) through the Eccellenza Professorial Fellowship ``Flavor Physics at the High Energy Frontier'' project number 186866. The work of AG and NS is also partially supported by the European Research Council (ERC) under the European Union’s Horizon 2020 research and innovation programme, grant agreement 833280 (FLAY). PN acknowledges the Humboldt foundation for support and the Max Planck Institute for Physics for hospitality. The work of LB is supported by the UZH Postdoc Grant Forschungskredit K-72324-03.

\appendix
\section{Instructions to run the code}
\label{app:App1}

The purpose of this appendix is to provide a brief guide to run the code. By the end of this section the reader should be able to reproduce the plots such as the ones in Figures~\ref{fig:mLQyLQ}, \ref{fig:dphiptlq}, \ref{fig:plotsjet} and~\ref{fig:plotslep}. 
The first step is to download the \POWHEGBOXRES{} and then get the process {\tt LQ-s-chan} from the svn repository svn://powhegbox.mib.infn.it/trunk/User-Processes-RES/LQ-s-chan. At the time of writing
\POWHEGBOXRES{} is at revision 3967.

 The Makefile 
 may need a few modifications. At the beginning choose the compiler and check that the commands to invoke the compiler ({\tt{F77}}, {\tt{CC}} and {\tt{CXX}}) match your system. On new MacOS  gcc and g++  by default point to clang and clang++. This can lead to problems when linking against libraries built with the actual GNU Compiler Collection (gcc).\footnote{It
is possible to run this on the new Apple Silicon processors if all packages are
built using homebrews gcc and gfortran.} LHAPDF is used to access the 
lepton PDFs {\tt{LUXlep-NNPDF31\_nlo\_as\_0118\_luxqed}}.  Therefore, the {\tt{lhapdf-config}} executable should be in the path. Set the variable {\tt{RES}} to the path of \POWHEGBOXRES{} following the examples in the file. Now it should be possible to build both targets ({\tt{pwhg\_main}} and {\tt{lhef\_analysis}}).

In order to shower the events one needs to download the appropriate version of \HERWIG{7}. In the folder
{\tt{HerwigInstallation}} one can find a simple installation script. It can be run directly, or used as
a sequence of instructions to install \HERWIG{7}.
The next step is to build the Herwig interface. To do so edit the Makefile in 
the folder {\tt{HerwigInterface}}. The variable {\tt{PROCDIR}} has to be set to the 
path of the LQ-Res-Prod folder. Again set the path to \POWHEGBOXRES{} and check whether the {\tt{herwig-config}} and {\tt{thepeg-config}} executables are in the path. Also, set the correct path to the {\tt{HepMC2}} library. After building the interface return to the project's main folder. Finally navigate to the folder's scripts and build the two executables 
{\tt{mergedata}} and {\tt{pastegnudata}}. Move them to a directory in the path. 
Everything needed to compute the histograms should now be compiled. 

To quickly check whether the code is yielding results open the script {\tt{run.sh}} and adjust the variables {\tt{ncores}} and {\tt{nprocesses}} to the system. To execute the code create two directories, one for the LO and one for the NLO computation. Copy the content of the folder {\tt{run-master}} to both folders. Now the input cards for \POWHEGBOXRES{} and Herwig should be present among some scripts to run the code on multiple cores. In the LO folder rename {\tt{powheg.input-save-LO}} to {\tt{powheg.input-save}}. Among many parameters that control the behaviour of \POWHEGBOXRES{} the mass and charge of the desired leptoquark is specified in this file. The mass and the charge of the leptoquark, as well as, the quark and lepton flavours, can be set. The latter is done by enabling the coupling for the corresponding family of quarks and leptons. The flavour of the quarks is determined by the charge of the leptoquark. If the number of events was changed in the POWHEG input card, the corresponding line in the Herwig input card {\tt{Herwig.in}} should be modified. 

To run the code, modify the lines controlling the number of cores and processes in the {\tt{run-parallel.sh}} script and execute it. This script will run multiple instances of \POWHEGBOXRES{} and create the Les Houches events files. For the analysis, execute the {\tt{runlhe.sh}} script. The parton  shower and its analysis is done by running the script {\tt{hw7.sh}}. Move the files with the top-extensions from the {\tt{HerwigRun}} directory up to the current directory and run {\tt{refine.sh}} combine the data from all processes. 
The same procedure can be repeated for  the NLO case. To plot the histograms create a new directory and copy the  python scripts to it. Set the variable {\tt{RUNDIRLO}} and {\tt{RUNDIRNLO}} to the directories containing the tables created with {\tt{refine.sh}}. Run the python script {\tt{plots.py}} to obtain the histograms.

\bibliography{LQ}
\bibliographystyle{JHEP}

\end{document}